\documentclass{aa}
\usepackage{graphicx}

\newcommand{\ratioo} {N({\rm H}_2) / I_{\rm CO}}

\begin{document}      

   \title{The influence of the cluster environment on the star formation efficiency of 12 Virgo spiral galaxies}

   \author{B.~Vollmer\inst{1}, O.I.~Wong\inst{2}, J.~Braine\inst{3,4}, A.~Chung\inst{5}, \and J.D.P.~Kenney\inst{6}}

   \offprints{B.~Vollmer, e-mail: Bernd.Vollmer@astro.unistra.fr}

   \institute{CDS, Observatoire astronomique, UMR 7550, 11, rue de l'universit\'e,
	      67000 Strasbourg, France \and
	      CSIRO Astronomy \& Space Science, Epping, NSW 1710, Australia \and
	      Univ. Bordeaux, Laboratoire d'Astrophysique de Bordeaux, UMR 5804, F-33270, Floirac, France \and
	      CNRS, LAB, UMR 5804, F-33270, Floirac, France \and
	      Department of Astronomy and Yonsei University Observatory, Yonsei University, Republic of Korea \and
	      Yale University Astronomy Department, P.O. Box 208101, New Haven, CT 06520-8101, USA
              }

   \date{Received / Accepted}

   \authorrunning{Vollmer et al.}
   \titlerunning{The influence of cluster environment on star formation efficiency}

\abstract{
The influence of the environment on gas surface density and star formation efficiency of cluster spiral galaxies is investigated.
We extend previous work on radial profiles by a pixel-to pixel analysis looking for asymmetries due to environmental interactions.
The star formation rate is derived from GALEX UV and Spitzer total infrared data based on the
8, 24, 70, and 160~$\mu$m data. 
As in field galaxies, the star formation rate for most Virgo galaxies is approximately proportional to the molecular gas mass. 
Except for NGC~4438, the cluster environment does not affect the star formation efficiency with respect to the molecular gas.
Gas truncation is not associated with major changes in the total gas surface density distribution of the inner disk of Virgo spiral galaxies.
In three galaxies (NGC~4430, NGC~4501, and NGC~4522), possible increases in the molecular fraction and the star formation efficiency
with respect to the total gas, of factors of $1.5$ to $2$, are observed on the windward side 
of the galactic disk.
A significant increase of the star formation efficiency with respect to the molecular gas content on the windward side of ram 
pressure-stripped galaxies is not observed.
The ram-pressure stripped extraplanar gas of 3 highly inclined spiral galaxies (NGC~4330, NGC~4438, and NGC~4522) shows a depressed star formation 
efficiency with respect to the total gas, and one of them (NGC~4438) shows a depressed rate even with respect to the molecular gas. 
The interpretation is that stripped gas loses the gravitational confinement and associated pressure of the galactic disk, and the gas flow is diverging, 
so the gas density decreases and the star formation rate drops. We found two such regions of low star formation efficiency 
in the more face-on galaxies NGC~4501 and NGC~4654 which are both undergoing ram pressure stripping. These regions show low radio continuum 
emission or unusually steep radio spectral index. However, the stripped extraplanar gas in one highly inclined galaxy (NGC~4569) shows a normal 
star formation efficiency with respect to the total gas. We propose this galaxy is different because it is observed long after peak pressure, 
and its extraplanar gas is now in a converging flow as it resettles back into the disk.
}

\maketitle

\section{Introduction \label{sec:intro}}

The cluster environment affects spiral galaxies in two main ways: (i) gravitational interactions between the galaxy and
the gravitational potential of the cluster or between two galaxies lead to distortions of the stellar and gas
distribution and (ii) the hydrodynamical interaction between the hot intracluster gas and the galaxy's ISM 
(ram pressure stripping) which does not act on the stars. The signposts of active ram pressure stripping are
(i) truncated gas disks within the optical radius and/or a one-sided gas tail together with a symmetric
stellar distribution (Chung et al. 2007, 2009), (ii) asymmetric ridges of polarized radio continuum emission (Vollmer et al. 2007)
and (iii) low radio-to-far infrared ratios (radio-deficit regions; Murphy et al. 2009) at the opposite side of the gas tail. 
Dynamical modelling of several Virgo cluster spiral galaxies lead to the following overall picture:
the extraplanar gas tails are of relatively high column density if
we observe the galaxy near a strong ram pressure peak. After peak ram pressure, the column density of the tail decreases 
significantly with time (Vollmer 2009). The polarized radio continuum emission traces the compression of the
ISM by ram pressure. In the compression region, the gas surface density is expected to be enhanced with respect
to that of an isolated symmetric galaxy. This may be actually observed in the best-studied case of a more face-on galaxy undergoing active
ram pressure stripping in the Virgo cluster, NGC~4501 (Vollmer et al. 2008a), where the H{\sc i} surface density
distribution is asymmetric, and higher on the windward side. The compression is expected to be stronger for small
inclination angles  between the galactic disk and the ram pressure wind (more edge-on stripping), since the gas near the 
leading edge is pushed into the gas of the inner disk. 

How does star formation react to the ram-pressure induced distortions of the gas distribution? Observationally, the observed
star formation rate $\dot{\Sigma}_{*}$ is linked to the total gas column density $\Sigma_{\rm g}$ by a power law (Schmidt law):
$\dot{\Sigma}_{*} \propto \Sigma_{\rm g}^{N}$ with $1.5 \leq N \leq 2$ (Schmidt 1963, Kennicutt 1998a). This law is widely used and mostly
based on integrated quantities of the molecular and atomic gas, and the star formation activity of a galaxy.
However, one should bear in mind that it is an oversimplification of reality (e.g., Wong \& Blitz 2002, Jogee et al. 2005, Wong 2009).
Recent investigations on resolved gas and star formation distributions of nearby spiral galaxies with $\sim 1$~kpc resolution 
show that beyond the central $\sim 1$~kpc star formation is approximately proportional to the molecular gas surface density 
($N \sim 1$; Bigiel et al. 2008, 2011). 
This means that the star formation efficiency of the molecular gas $SFE_{\rm mol}=\dot{\Sigma}_{*}/\Sigma_{\rm H_{2}}$ is constant
with a $1\sigma$ scatter of 0.24~dex (Bigiel et al. 2011).

Most galaxies show little or 
no pixel-to-pixel correlation between star formation and the atomic gas surface density (Bigiel et al. 2008). 
On the other hand, the radial profiles of the total gas surface density are as well correlated to the star formation rate as those 
of the molecular gas (Leroy et al. 2008, Schruba et al. 2011). According to the latter authors, both relationships have roughly the same
rank correlation coefficient.
Whereas the SFR-H$_{2}$ correlation is in most cases linear, the SFR-(H{\sc i}+H$_{2}$) shows a break at a gas surface density of 
$\sim 14$~M$_{\odot}$pc$^{-2}$.
The slope of the SFR-(H{\sc i}+H$_{2}$) relation in the H{\sc i} dominated outer regions ($R > 0.4 R_{25}$) is much steeper than that 
of the inner region ($R < 0.4 R_{25}$) dominated by the molecular gas.

Do the star formation correlations, which were derived from nearby field spiral galaxies, still hold in perturbed Virgo cluster galaxies?
Fumagalli et al. (2008) used Nobeyama 45m (Kuno et al. 2007), VIVA (VLA Imaging of Virgo in Atomic gas) H{\sc i} (Chung et al. 2009), and 
H$\alpha$ imaging of 10 Virgo spiral galaxies to study the relation between the radial profiles of gas content and star formation.
They showed that the bulk of star formation correlates with the molecular gas, but the atomic gas is claimed to be important in supporting 
the star formation activity in the outer part of the disks.

We extend the work of Fumagalli et al. (2008) presenting a sample of 12 Virgo spiral galaxies (Table~\ref{tab:table}) 
for which imaging H{\sc i}, CO, infrared, and UV data exist. We are interested in the gas content and star formation
efficiency on a pixel-by-pixel basis to detect asymmetries due to environmental interactions. Our results are compared to
the recent work based on THINGS (Walter et al. 2008) and HERACLES (Leroy et al. 2009) data.

This article is structured in the following way: the data, on which the gas surface density and the star formation rate
are based, are described in Sect.~\ref{sec:data}. We also provide a comparison between the total infrared emission estimate based
on the 8, 24, 70, and 160~$\mu$m and that based on the 24~$\mu$m data alone. The results for the images of the
atomic/molecular/total gas surface density and star formation rate are presented galaxy by galaxy in Sect.~\ref{sec:results}.
These results are discussed and interrelated in Sect.~\ref{sec:discussion} followed by our conclusions (Sect.~\ref{sec:conclusions}).

\section{Data \label{sec:data}}

We derive our sample of Virgo galaxies from the VIVA--SPITSOV (SPITzer Survey Of Virgo)
multiwavelength surveys of 44 Virgo spirals which span a range of gas, stellar mass 
and star formation properties spread throughout the cluster.  The VLA Imaging of Virgo in Atomic gas (VIVA; Chung et al 2009) provide the H{\sc i}
observations, while the mid-infrared (MIR) to far-infrared (FIR) imaging are obtained from the Spitzer Survey of Virgo (SPITSOV; Kenney et al. 
2009, Wong et al. 2012a in prep.). The observations from SPITSOV are more sensitive by a factor of $\sqrt{2}$  than the SINGS observations 
as SPITSOV is largely interested in detecting extraplanar and outer galaxy IR emission (Wong et al. 2012a in prep.). Our large multiwavelength 
ancillary database also includes optical $BVR$ and H$\alpha$ imaging (Koopmann \& Kenney 2004) as well as optical spectroscopy (Crowl \& Kenney 2008).
Furthermore, we use H$\alpha$ images of NGC~4522 from Kenney et al. (2004), NGC~4330 from Abramson et al. (2011), and
NGC~4438 from GOLDMine (Gavazzi et al. 2003).

The gas surface densities are determined from the HI VIVA observations as well as CO observations from Nobeyama 45m/ Pico Veleta 30m. 
We estimate the star formation rates based on ultraviolet (UV) observations from the GALEX space telescope (Gil de Paz et al. 2007) in 
combination with the SPITSOV observations. Compared to Fumagalli et al. (2008) we added NGC~4330, NGC~4438, and NGC~4522 to the sample.
NGC~4330 and NGC~4522 are undergoing active ram pressure and show extraplanar H{\sc i} and UV emission (Abramson et al. 2011,
Kenney et al. 2004). NGC~4438 had a strong tidal interaction $\sim 100$~Myr ago and is now undergoing strong active ram pressure stripping
(Vollmer et al. 2005b).
We removed NGC~4535 from the sample, because Spitzer data are not available for this galaxy. Our Virgo galaxy sample is
presented in Table~\ref{tab:table}.

\begin{table*}
      \caption{Spatial resolutions, gas morphologies, and interaction types.}
         \label{tab:table}
      \[
         \begin{array}{lcccccccc}
           \hline
           \noalign{\smallskip}
           {\rm galaxy\ name } & {\rm m_{\rm B}} & i^{(1)}  & {\rm D_{25}}  & {\rm HI\ linewidth}^{(2)} & {\rm Dist.^{(3)}} & {\rm resolution^{(4)}} & {\rm \ \ \ \ \ gas\ distribution\ \ \ \ \ } & {\rm interaction\ type} \\
	    & {\rm (mag)} & {\rm (deg)} & {\rm arcmin}  & {\rm (km/s)} & {\rm (deg)} & {\rm (arcsec)} & & \\
	   \noalign{\smallskip}
	   \hline
	   \noalign{\smallskip}
	   {\bf NGC~4254} & 10.44 & 30 & 5.37 & 250 & 3.6 & 40 & {\rm normal\ +\ HI\ tail} & {\rm tidal} \\ 
           \noalign{\smallskip}
	   \hline
	   \noalign{\smallskip}
	   {\bf NGC~4321} & 10.02 & 27 & 7.41 & 268 & 4.0 & 32 & {\rm normal} & {\rm none} \\ 
           \noalign{\smallskip}
	   \hline
	   \noalign{\smallskip}
	   {\bf NGC~4330} & 13.09 & 84 & 4.47 & 275 & 2.1 & 28 & {\rm truncated\ +\ HI\ tail} & {\rm active\ ram\ pressure} \\
	   \noalign{\smallskip}
           \hline
	   \noalign{\smallskip}
	   {\bf NGC~4402} & 12.64 & 80 & 3.89 & 288 & 1.4 & 18 & {\rm truncated\ +\ HI\ tail}  & {\rm active\ ram\ pressure} \\
	   \noalign{\smallskip}
           \hline
	   \noalign{\smallskip}
	   {\bf NGC~4419} & 12.08 & 74 & 3.31 & 382 & 2.8 & 18 & {\rm anemic\ +\ truncated} & {\rm past\ ram\ pressure?} \\
	   \noalign{\smallskip}
           \hline
	   \noalign{\smallskip}
	   {\bf NGC~4438} & 10.39 & \sim 70 & 9.3 &  & 1.0 & 25 & {\rm truncated+extraplanar\ gas} & {\rm tidal + active\ ram\ pr} \\
	   \noalign{\smallskip}
           \hline
	   \noalign{\smallskip}
	   {\bf NGC~4501} & 10.50 & 57 & 6.92 & 532 & 2.0 & 18 & {\rm truncated} & {\rm active\ ram\ pressure} \\
	   \noalign{\smallskip}
           \hline
	   \noalign{\smallskip}
	   {\bf NGC~4522} & 12.99 & 79 & 3.72 & 240 & 3.3 & 20 & {\rm truncated+extraplanar\ gas} & {\rm active\ ram\ pressure}\\
	   \noalign{\smallskip}
           \hline
	   \noalign{\smallskip}
	   {\bf NGC~4548} & 10.96 & 38 & 5.37 & 249 & 2.4 & 18 & {\rm anemic} &  {\rm active\ ram\ pressure?} \\
	   \noalign{\smallskip}
           \hline
	   \noalign{\smallskip}
	   {\bf NGC~4569} & 10.26 & 65 & 9.55 & 406 & 1.7 & 18 & {\rm truncated} & {\rm past\ ram\ pressure} \\
	    \noalign{\smallskip}
           \hline
	   \noalign{\smallskip}
	   {\bf NGC~4579} & 10.48  & 38 & 5.89 & 371 & 1.8 & 44 & {\rm anemic\ +\ truncated} & {\rm none?}\\
	   \noalign{\smallskip}
           \hline
	   \noalign{\smallskip}
	   {\bf NGC~4654} & 11.31 & 51 & 4.90 & 310 & 3.4 & 18 & {\rm normal\ +\ HI\ tail} & {\rm tidal + active\ ram\ pr} \\
	   \noalign{\smallskip}
           \hline
        \end{array}
      \]
\begin{list}{}{}
\item[$^{(1)}$Inclination angle; $^{(2)}$observed HI width at $20$\,\% (from Chung et al. 2009);  $^{(3)}$Distance from M~87; $^{(4)}$common image resolution] 
\end{list}
\end{table*}

\subsection{Gas surface density \label{sec:gasdens}}

We calculate the local atomic mass surface density, $\Sigma_{\rm HI}$, from VIVA H{\sc i} observations (Chung et al. 2009).
To convert from integrated intensity to $\Sigma_{\rm HI}$, we use 
\begin{equation}
\Sigma_{\rm HI}\ {\rm (M_{\odot}pc^{-2})}=0.020 \times I_{\rm 21cm}\ {\rm (K\ km\,s^{-1})}\ ,
\end{equation}
which includes a factor of 1.36 to reflect the presence of helium (see, e.g., Leroy et al. 2008).

The surface density of the molecular hydrogen, $\Sigma_{\rm H_{2}}$ is estimated from CO emission. Following Leroy et al. (2008) we derive
the $\Sigma_{\rm H_{2}}$ from integrated CO intensity, $I_{\rm CO}$, by adopting a constant CO-to-H$_2$ conversion factor
X$_{\rm CO} = 2 \times 10^{20}$~cm$^{-2}$(K\ km\,s$^{-1}$)$^{-1}$ and accounting for helium. This yields
\begin{equation}
\Sigma_{\rm H_{2}}\ {\rm (M_{\odot}pc^{-2})}=4.4 \times I_{\rm CO}(1-0)\ {\rm (K\ km\,s^{-1})}\ .
\end{equation}
If CO(2-1) observations are available, we assume $I_{\rm CO}(2-1)=0.8 \times I_{\rm CO}(1-0)$.
All surface densities are projected surface densities.

Nobeyama 45m telescope BEARS CO(1--0) data cubes (Kuno et al. 2007) were used for all galaxies except NGC~4330, NGC~4438, and NGC~4522, 
for which IRAM 30m HERA CO(2--1) observations exist and are used as presented in Vollmer et al. (2008b, 2011). 
For all Nobeyama CO data, we produced maps of the
gas distribution (moment~0) by taking advantage of the H{\sc i} data cubes assuming that the CO line lies within
the H{\sc i} line profile (see Vollmer et al. 2008). We first checked that the CO lines, where clearly visible, were 
located within the H{\sc i} line profiles.
A first 3D binary mask was produced by clipping the H{\sc i} data cube at the $3\sigma$ level.
For the CO data cube, we calculated the rms noise level for each spectrum in the velocity regions
devoid of an H{\sc i} signal and subtracted a constant baseline. A second 3D binary mask was produced by clipping the CO data cube 
at a $3$-$4\sigma$ level according to the data quality.  
The H{\sc i} and CO 3D binary masks were added, applied to the CO data cube, and moment~0 and moment~1 maps were created.
If the moment~0 map did not show residual noise and the moment~1 map did not contain anomalies, we stopped the 
CO data cube processing (NGC~4321, NGC~4419, NGC~4501, NGC~4569, NGC~4579). In the other cases (NGC~4254, NGC~4402, NGC~4548, NGC~4654), 
a polynomial of order $3$ was fitted to the CO spectrum in the velocity regions devoid of an H{\sc i} signal and subtracted from the spectra.
For these four galaxies, the second 3D binary mask was reconstructed by clipping the CO data cube requiring a $3\sigma$ level in three adjacent 
velocity channels. 
The H{\sc i} and CO 3D binary masks were then multiplied, the resulting mask was applied to the CO data cube, and moment~0 and 
moment~1 maps were created. Because the signal-to-noise ratio of the Nobeyama CO data of NGC~4548, which displays a central H{\sc i} hole, 
did not permit the detection of  the central region including the bar, we added deeper IRAM 30m CO(1--0) map of this central part of the 
galaxy (Vollmer et al. 1999) to the Nobeyama CO map.
The typical rms noise levels were $20$-$30$~mK for both processing options. A $3\sigma$ detection in
three adjacent velocity channels of $5$~km\,s$^{-1}$ leads to a limiting H$_{2}$ surface density of $4$-$6$~M$_{\odot}$pc$^{-2}$. 
This procedure thus allowed for deeper and more reliable CO maps.
To demonstrate the improvement of our new CO distribution maps compared to those resulting from clipping 
the CO data cube at a constant level of $75$~mK which corresponds to the mean rms noise level in a $5$~km\,s$^{-1}$ channel of the
Nobeyama CO atlas of nearby spiral galaxies given by Kuno et al. (2007), we show both maps for NGC~4501 and NGC~4402 in 
Fig.~\ref{fig:compareCO}.

The spatial resolution of the CO(1--0) observations is $15''$, that of the CO(2--1) $11''$.
The resolution of the H{\sc i} maps varies between $18''$ and $40''$ (Table~\ref{tab:table}).
For the combination of both maps, we convolved the CO data to the H{\sc i} resolution.
The total gas surface density was then calculated as $\Sigma_{\rm g} = \Sigma_{\rm H_{2}} + \Sigma_{\rm HI}$.

\subsection{Star formation rate}

The star formation rate was calculated from the FUV luminosities corrected by the total infrared to FUV luminosity ratio (Hao et al. 2011).
This method takes into account the UV photons from young massive stars which escape the galaxy and those which
are absorbed by dust and re-radiated in the far infrared:
\begin{equation}
\dot{\Sigma}_{*} = 8.1 \times 10^{-2}\ (I({\rm FUV}) + 0.46\ I({\rm TIR}))\ ,
\end{equation}
where $I({\rm FUV})$ is the GALEX far ultraviolet and $I({\rm TIR})$ the total infrared intensity based on Spitzer IRAC and MIPS data
in units of MJy\,sr$^{-1}$. $\dot{\Sigma}_{*}$ has the units of M$_{\odot}$kpc$^{-2}$yr$^{-1}$.
This prescription only holds for a constant star formation rate over the last few 100~Myr.
Since the star formation rate of Virgo galaxies can change more rapidly than this timescale due to environmental effects in the cluster
(as in NGC~4330, NGC~4438, and NGC~4522), the uncertainty on the derived star formation rate becomes larger in these cases.
For these three galaxies we also show that star formation rate distributions based on H$\alpha$, 
which traces recent star formation within the last $10$~Myr,
and TIR (Figs.~\ref{fig:plot_n4330_ha}, \ref{fig:plot_n4438_ha}, and \ref{fig:plot_n4522_ha}).  
For NGC~4330, we used the total H$\alpha$ flux from Goldmine (Gavazzi et al. 2003), for NGC~4522 that of Kenney et al. (2004) to calibrate the
WIYN H$\alpha$ images. For NGC~4438, the Goldmine H$\alpha$ flux calibration was adopted. We derived the star formation rate from the H$\alpha$ 
surface brightness following Kennicutt (1998b):
\begin{equation}
\dot{M}_{*} = L(H_{\alpha})/ (1.26 \times 10^{41}~{\rm erg\,s}^{-1})\ ({\rm M}_{\odot}{\rm yr}^{-1}).
\end{equation}

The full width at half-maximum (FWHM) of the point spread functions (PSFs), as stated in the Spitzer Observer's Manual 
(Spitzer Science Centre 2006), are 1.7, 2.0, 6, 18 and 38~arcsec at 3.6, 8.0, 24, 70 and 160~$\mu$m, respectively.
First, the data are convolved with kernels that match the PSFs of the images in the 3.6, 8, 24 and 70~$\mu$m bands to the PSF of 
the H{\sc i} data (Table~\ref{tab:table}). For all galaxies, except NGC~4254 and NGC~4579, the H{\sc i} resolution is
significantly higher than that of the 160~$\mu$m data. Since we want to keep the highest possible resolution ($\sim 20''$),
the 160~$\mu$m data were only included for NGC~4254 and NGC~4579 in the analysis.
Next, the data were re-binned to the common pixel size of the VIVA H{\sc i} maps. 
We exclude from the analysis regions not detected at the $3 \sigma$ level in one or more wave bands.
This resulted in a loss of surface of $10$\,\% to $20$\,\% compared to the regions with $24~\mu$m $3\sigma$ detections
with a tendency of less loss for truncated galaxies.
Following Helou et al. (2004), we subtracted the stellar continuum from the 8 and 24~$\mu$m surface brightnesses (in MJy\,sr$^{-1}$) using 
\begin{equation}
I_{\nu}({\rm PAH}\ 8\mu {\rm m})=I_{\nu}(8\mu {\rm m})-0.232\ I_{\nu}(3.6\mu {\rm m})\ {\rm and}
\end{equation}
\begin{equation}
I_{\nu}(24\mu {\rm m})=I_{\nu}(24\mu {\rm m})-0.032\ I_{\nu}(3.6\mu {\rm m})\ .
\end{equation}
We calculated the TIR surface brightnesses using
\begin{eqnarray}
I({\rm TIR})=0.95 \nu I_{\nu}({\rm PAH}\ 8\mu {\rm m}) + 1.15 \nu I_{\nu}(24\mu {\rm m}) + \nonumber \\
\nu I_{\nu}(70\mu {\rm m}) + \nu I_{\nu}(160\mu {\rm m})
\label{eq:tir}
\end{eqnarray}
based on equation (22) from Draine \& Li (2007; see also Bendo et al. 2008).
If the H{\sc i} image resolution is smaller than $30''$, we did not want to degrade the resolution of the final image set  
to the $38''$ resolution of the $160$~$\mu$m data. In these cases we replaced $I_{\nu}(160\mu {\rm m})$ by $2.3 \times I_{\nu}(70\mu {\rm m})$,
which corresponds to the observed relation for NGC~4254 and NGC~4579.
In most galactic environments the error introduced to $I({\rm TIR})$ is about 50\,\% based on the Draine \& Li models.

\begin{figure*}
  \centering
  \includegraphics[width=16cm]{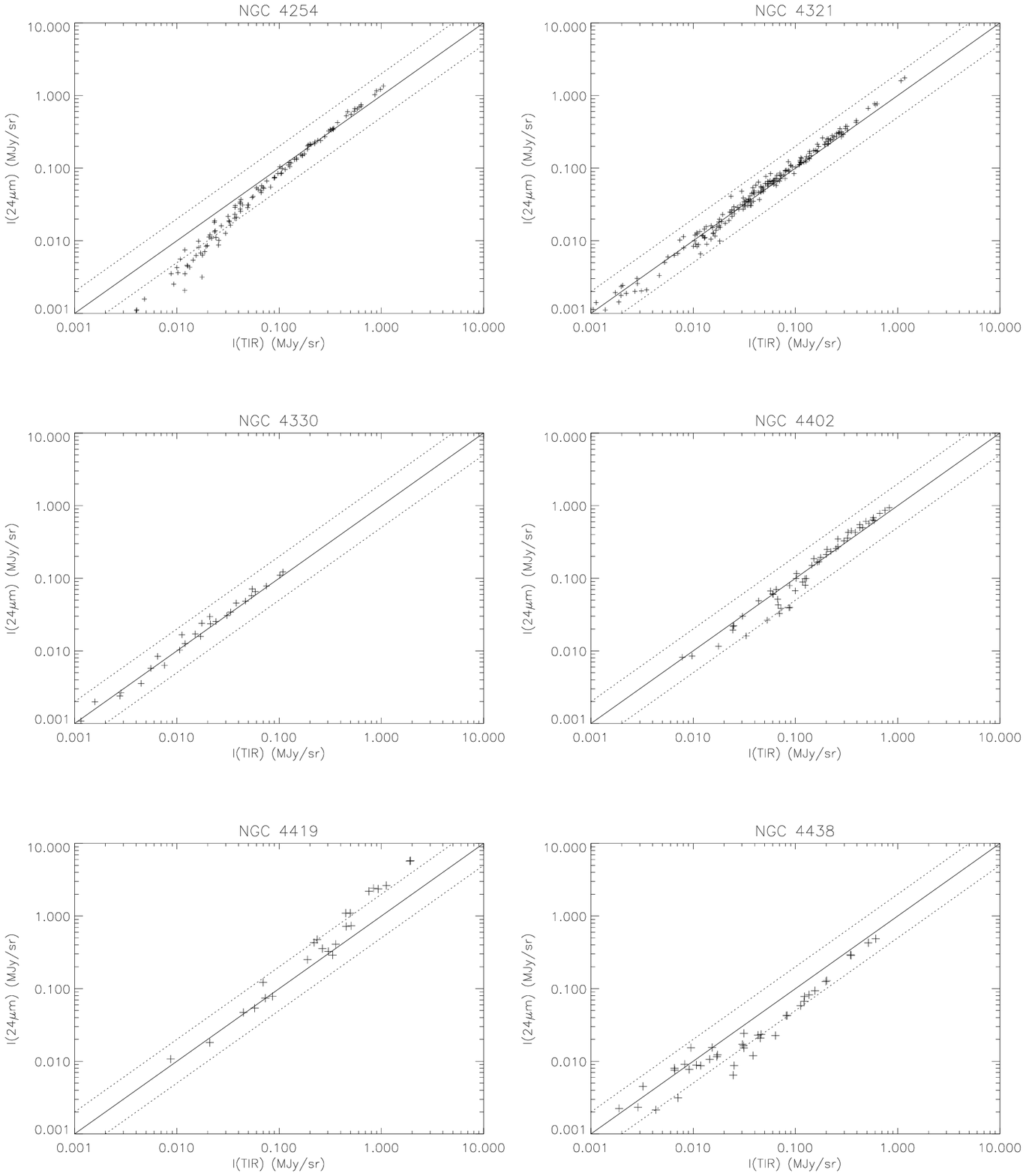}
  \caption{Spitzer $24$~$\mu$m intensity $I(24\mu {\rm m})$ as a function of the total infrared
    intensity $I({\rm TIR})$ calculated using Eq.~\ref{eq:tir} on a pixel-by-pixel basis.
      The pixel have the size of a resolution element. The dotted lines correspond to offsets of 0.3~dex.
  \label{fig:ir_1}}%
\end{figure*}

\begin{figure*}
  \centering
  \includegraphics[width=16cm]{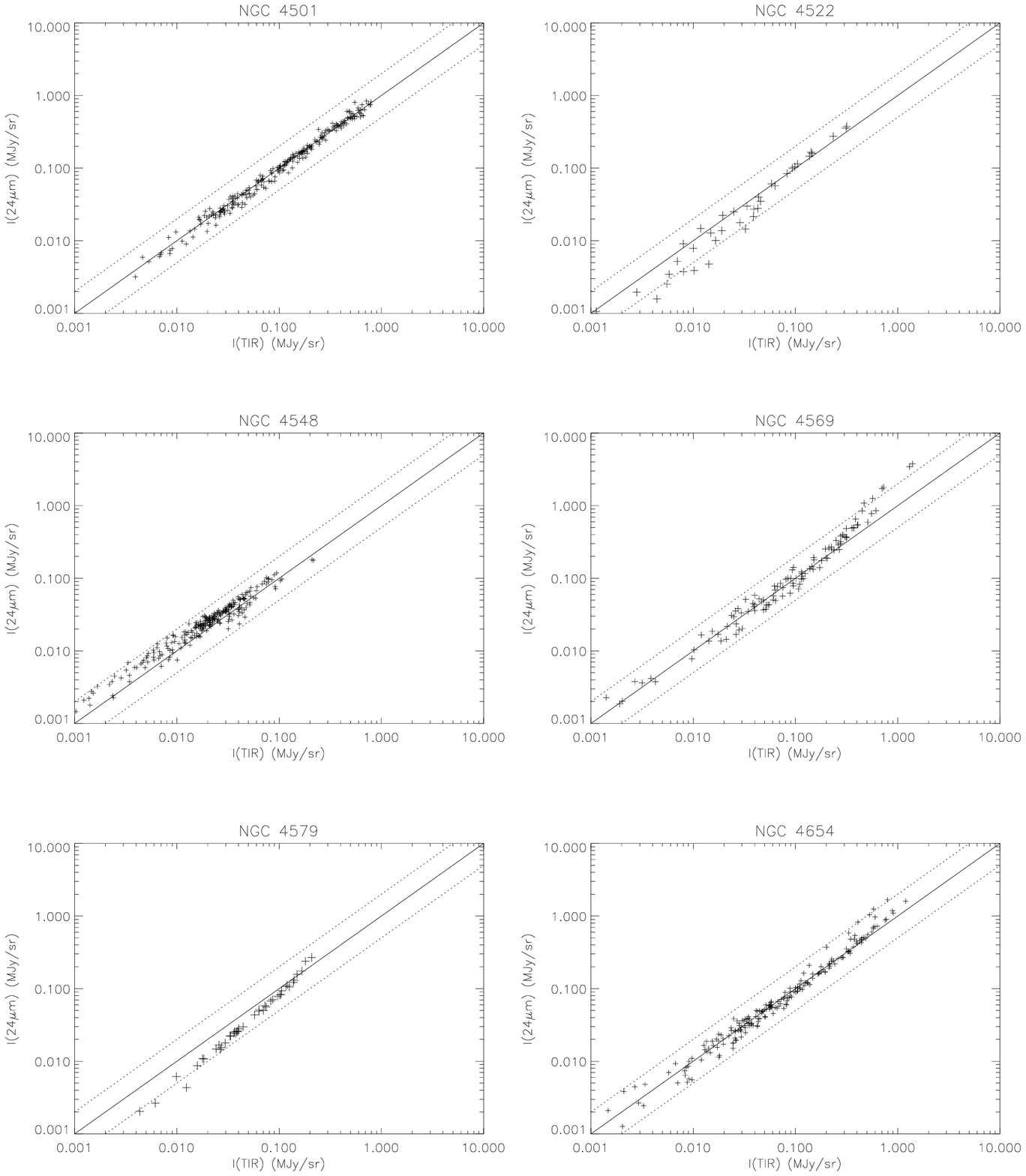}
  \caption{Spitzer $24$~$\mu$m intensity $I(24\mu {\rm m})$ as a function of the total infrared
    intensity $I({\rm TIR})$ calculated using Eq.~\ref{eq:tir} on a pixel-by-pixel basis.
      The pixel have the size of a resolution element. The dotted lines correspond to offsets of 0.3~dex.
  \label{fig:ir_2}}%
\end{figure*}

Leroy et al. (2008) and Bigiel et al. (2008) used a star formation recipe based only on the $24$~$\mu$m data:
\begin{equation}
\dot{\Sigma}_{*} =8.1 \times 10^{-2}\ I({\rm FUV}) + 3.2 \times 10^{-3} I(24\mu {\rm m})\ .
\label{eq:24mu}
\end{equation}
To check the consistency between the two approaches (Eq.~\ref{eq:tir} and \ref{eq:24mu}), we present $I(24\mu {\rm m})$
as a function of $I({\rm TIR})$ in Fig.~\ref{fig:ir_1} and \ref{fig:ir_2}.
As shown by Kennicutt et al. (2009) there is good agreement between the two approaches.
Since we included the $160$~$\mu$m data in the analysis of NGC~4254 and NGC~4579, we find that the TIR intensity based solely 
on the $24$~$\mu$m intensity underestimates the TIR intensity based on
Eq.~\ref{eq:tir} by up to a factor of two for low intensities around $0.01$~MJy\,sr$^{-1}$.
This is understandable, because the stellar radiation field in these regions is weak, the dust temperature 
is thus low, and the fraction of the $160$~$\mu$m emission to the TIR emission is high.
For the other galaxies where we approximated the $160$~$\mu$m intensity by $2.3$ times the $70$~$\mu$m intensity,
the TIR intensity based on the $24$~$\mu$m intensity follows closely the TIR intensity
based on Eq.~\ref{eq:tir} except in the galaxy centers where TIR intensity based on the $24$~$\mu$m intensity 
overestimates that based on Eq.~\ref{eq:tir}. This overestimation is most prominent for
the nuclei of NGC~4419 and NGC~4569. 
We conclude that our star formation rate estimates are robust, with maximum uncertainties of up to a factor of two 
in regions of high ($\sim 1$~MJy\,sr$^{-1}$) and low ($\sim 0.01$~MJy\,sr$^{-1}$) infrared surface brightnesses.

\subsection {Caveats on the H$_2$ mass and SFR}

As in other works on the subject, we are using CO emission to trace the molecular hydrogen column density, used to calculate the SFE.

However, the amount of H$_2$ per unit CO emission, a.k.a. the $\ratioo$ ratio, clearly varies with galactocentric distance 
(Sodroski et al. 1995, Braine et al. 1997, and Braine et al. 2010). The $\ratioo$
ratio also increases with decreasing metallicity such that metallicity gradients may be at least partially responsible for the 
radial $\ratioo$ gradient in galaxies.  While the absolute calibration of metallicity measurements is uncertain, Magrini et al. (2011)
show that for at least several galaxies in our sample, all the calibrations indicate metallicity gradients.  Thus, it is likely that 
the H$_2$ mass in the outer parts of these galaxies is underestimated by the CO measurements and the use of a constant "standard" 
$\ratioo$ value chosen for compatibility with other studies.

The other ingredient in the SFE calculation, the star formation rate, is also affected by the metal content and probably other 
properties of the ISM.  The dust extinction, and therefore the fraction of the stellar energy absorbed and re-radiated by the 
dust, decreases with decreasing metallicity.  An opposite effect affects H$\alpha$ or Far-UV based estimates of the SFR as the 
emission in these bands is related to the energy $not$ absorbed by the dust.  This can be readily seen through the sharp gradient 
in the H$\alpha$/24 micron flux ratio in spirals (see e.g. Fig. 9 of Gardan et al. 2007), providing a justification for 
the use of a dust$+$H$\alpha$ or FUV based SFR indicator.  These effects should become more visible as measurement quality increases.
Moreover, the calibration of star formation rate indicators have been done for face-on galaxies and there may exist inclination effects 
which will contribute to the uncertainties.

\section{Results \label{sec:results}}

\begin{figure}
  \centering
   \resizebox{\hsize}{!}{\includegraphics{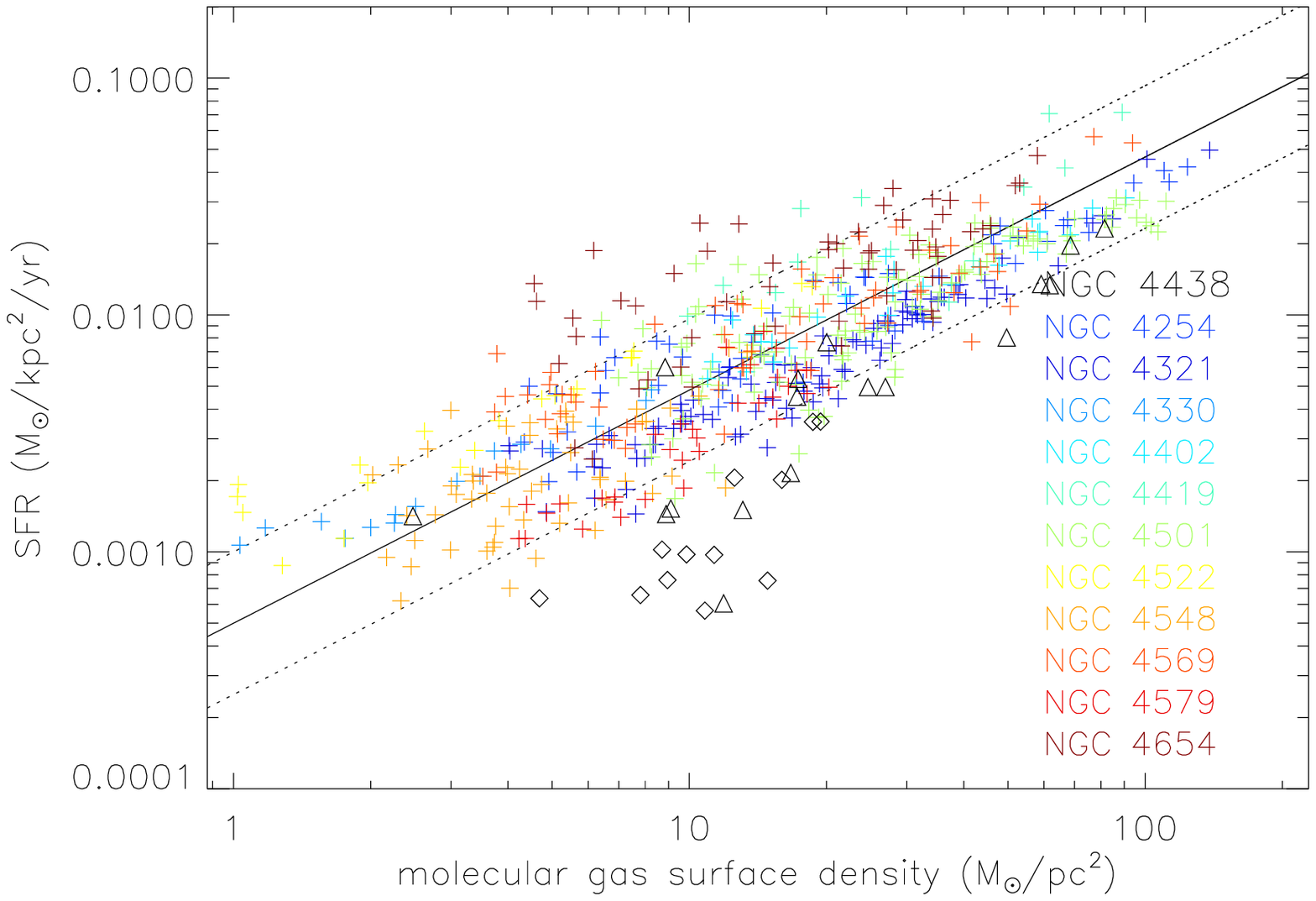}}
   \resizebox{\hsize}{!}{\includegraphics{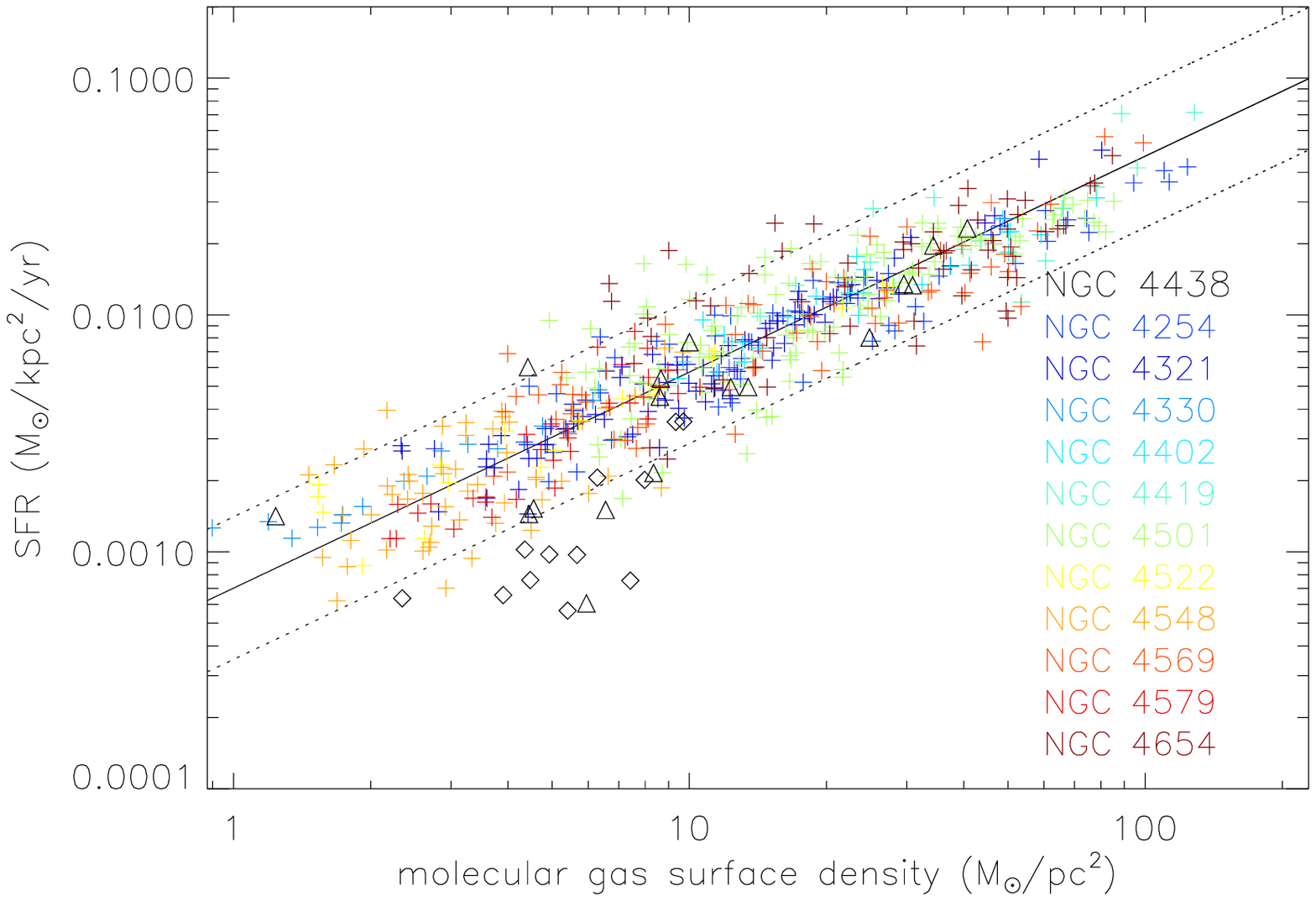}}
  \caption{Star formation rate as a function of the molecular gas surface density for all sample galaxies.
    Upper panel: data with a constant CO-to-H$_{2}$ conversion factor. Lower panel: with normalized molecular gas surface densities (see text). 
    The solid lines represent a linear regression using
    a Bayesian approach to linear regression with errors in both variables (Kelly 2007). To guide the eye, the dotted lines
    represent offsets of a factor of two. Triangles represent data points within the disk of NGC~4438 (RA offset $> -20''$),
    diamonds points in the western extraplanar region (RA offset $\leq -20''$; see Fig.~\ref{fig:n4438_sft_si}).
  \label{fig:sfr}}
\end{figure}

The resulting maps of the total gas surface density $\Sigma_{\rm g}=\Sigma_{\rm H_{2}}+\Sigma_{\rm HI}$, star formation rate
$\dot{\Sigma}_{*}$, molecular fraction $\Sigma_{\rm H_{2}}/\Sigma_{\rm HI}$, and star formation timescale 
$t_{*}=\Sigma_{\rm g}/\dot{\Sigma}_{*}$ for our sample galaxies are presented in the appendix. 
Also shown is the relationship between the molecular/atomic/total gas surface density and the star formation rate on a pixel-by-pixel basis.
The star formation rate as a function of the molecular gas surface density for all sample galaxies is shown in the upper panel of
Fig.~\ref{fig:sfr}. As Schruba et al. (2011), we observe systematic galaxy-to-galaxy variations of the star formation efficiency
with respect to the molecular gas $SFE_{\rm mol}$ of the order of $0.3$~dex. These SFR/L(CO) galaxy-to-galaxy variations
could be due to variations in the CO-to-H$_{2}$ conversion factor, or they could be intrinsic variations of $SFE_{\rm mol}$. 
Since we suspect variations in the CO-to-H$_{2}$ conversion factor may be at least partially responsible for the galaxy-to-galaxy SFR/L(CO) variations, 
we scale the molecular gas surface densities by a factor which normalizes the SFR/L(CO) values for each galaxy, following Schruba et al. (2011).
For NGC~4438, the normalization is based on the disk values. NGC~4522 and NGC~4654 show the highest normalization
($1.5$), NGC~4579 and NGC~4321 the lowest normalization ($0.5$ and $0.6$, respectively).
This normalization increased the Spearman rank correlation from $0.86$ to $0.92$.
The compilation of the normalized radial $SFE_{\rm mol}$ distributions (lower panel of Fig.~\ref{fig:sfr}) show the typical or average  
$SFE_{\rm mol}$ variations within a galaxy.
We performed a linear regression to the initial and normalized data with a method that uses a Bayesian approach to linear regression with 
errors in both variables (Kelly 2007).
A constant error of $0.3$~dex for the molecular gas surface density and $0.2$~dex for the star formation rate was assumed. The resulting slopes are 
$0.98 \pm 0.04$ for the data with a constant CO-to-H$_{2}$ conversion factor and 
$0.91 \pm 0.03$ for the normalized data (Fig.~\ref{fig:sfr}), i.e. the relation is very close to linear.
The associated average star formation efficiency of the data with a constant CO-to-H$_{2}$ conversion factor 
is $SFE_{\rm mol} = (5.5 \pm 3.6) \times 10^{-10}$~yr$^{-1}$.
The scatter of the normalized data is $3.1 \times 10^{-10}$~yr$^{-1}$.
Previous studies showed that the star formation efficiency with respect to the molecular gas ($SFE_{\rm mol}=\dot{\Sigma}_{*}/\Sigma_{H_{2}}$)
outside the central $\sim$kpc and averaged over regions $>1$~kpc is about constant in nearby spiral galaxies 
$SFE_{\rm mol} \sim 5 \times 10^{-10}$~yr$^{-1}$ with a $1\sigma$ scatter of 0.24~dex (Bigiel et al. 2008, Leroy et al. 2008). 
This is also the case for the galaxies in our sample, except for the extraplanar (RA offset $< -20''$) molecular gas of NGC~4438 (see Sect.~\ref{sec:n4438}).
The pixel-by-pixel correlation between the star formation rate and the atomic gas surface density is very weak (Bigiel et al. 2008). On the other
hand, the correlation between the radial profiles of the star formation rate and the total gas surface density ($SFE_{\rm tot}=\dot{\Sigma}_{*}/\Sigma_{\rm g}$,
where $\Sigma_{\rm g}=\Sigma_{\rm HI}+\Sigma_{\rm H_{2}}$, can be reasonably described by a broken power law (Leroy et al. 2008).
We also observe a change of the slope at $\Sigma_{\rm g} \sim 30$~M$_{\odot}$pc$^{-2}$.
In regions that are dominated by the molecular gas phase ($\Sigma_{\rm g} > 30$~M$_{\odot}$pc$^{-2}$), the pixel-by-pixel SFR-(H{\sc i}+H$_{2}$) 
correlation is linear, because $SFE_{\rm mol}$ is about constant. The correlation of the star formation rate with the
total gas surface density of our galaxy sample is less tight (Spearman rank correlation of $0.83$) than that with the molecular gas surface density
(Spearman rank correlation of $0.92$).
The use of the map of star formation efficiency with respect to the total gas $SFE_{\rm tot}$ is thus of limited use, because locally
it can take values between zero and $\sim 10^{-9}$~yr$^{-1}$. However, we think that it is still valuable to detect $SFE_{\rm tot}$ asymmetries 
caused by environmental interactions. 

\begin{figure}
  \centering
   \resizebox{\hsize}{!}{\includegraphics{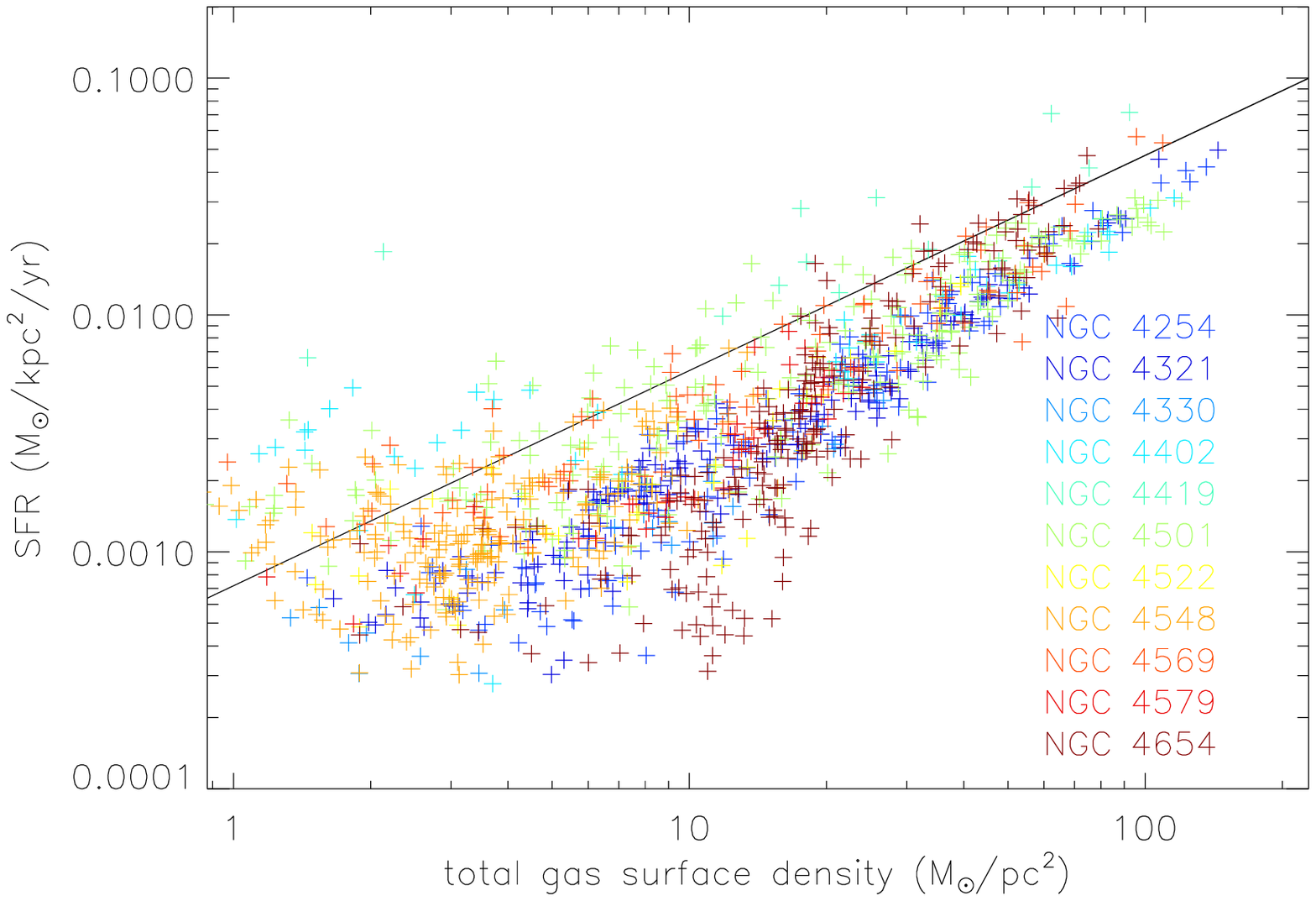}}
   \resizebox{\hsize}{!}{\includegraphics{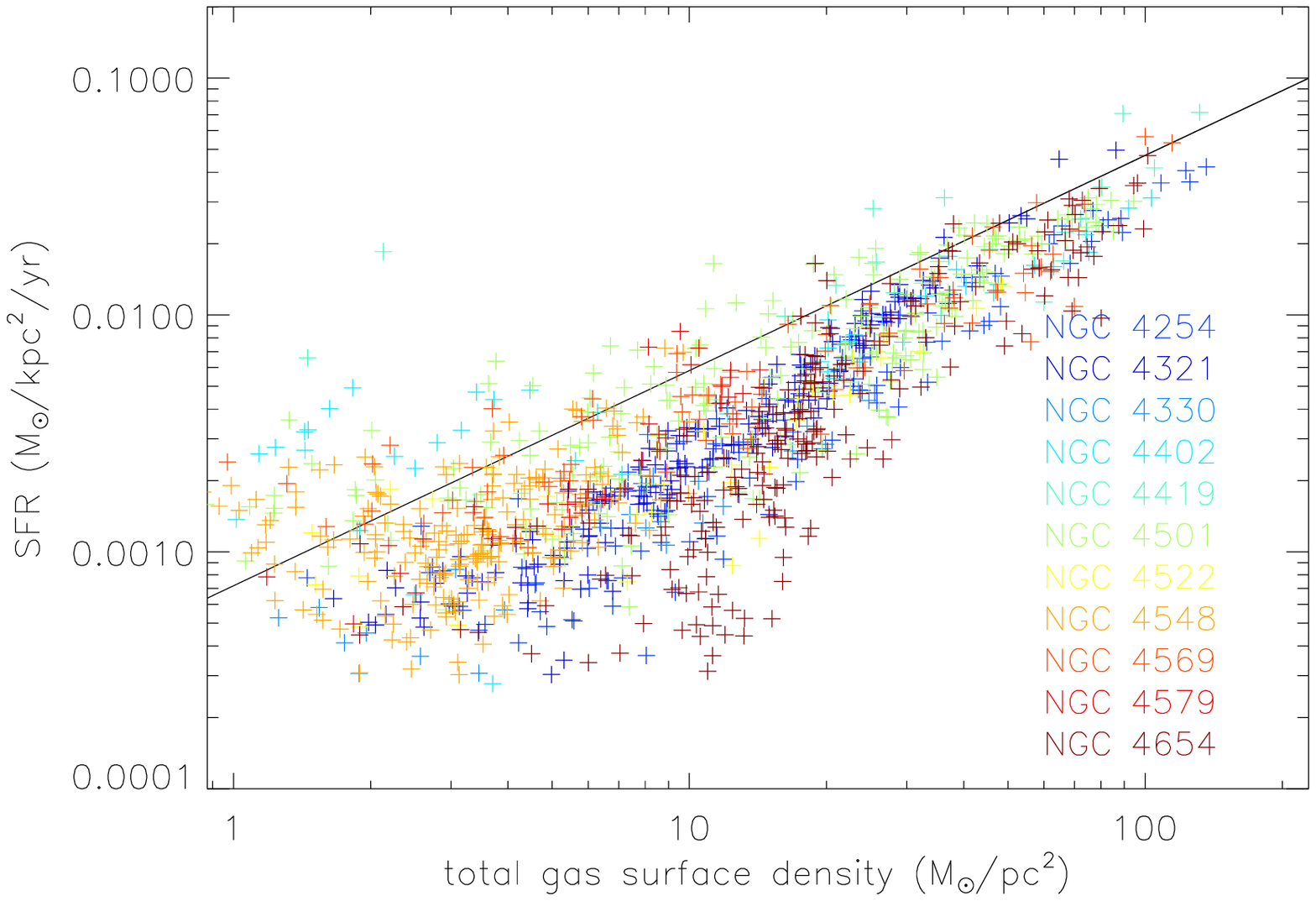}}
  \caption{Star formation rate as a function of the total gas surface density for all sample galaxies, except NGC~4438.
    Upper panel: data with a constant CO-to-H$_{2}$ conversion factor. Lower panel: with normalized molecular gas surface densities (see text). 
    The solid lines are the same as in Fig.~\ref{fig:sfr}.
  \label{fig:sfrtot}}
\end{figure}

In the following we comment on each galaxy. We first give additional information on environmental
interactions of the galaxies if available and then present our results. These are then correlated in Sect.~\ref{sec:discussion}.

NGC~4254 (Fig.~\ref{fig:plot_n4254}): this galaxy shows a huge ($\sim 100$~kpc) H{\sc i} plume to the north (Minchin et al. 2007, Haynes et al. 2007) 
most probably due to a past rapid flyby of a massive galaxy (Duc \& Bournaud 2008) which induced a prominent m=1 spiral arm structure. 
Ram pressure might also act on this galaxy from the southeast (Vollmer et al. 2005a, Kantharia et al. 2008, Murphy et al. 2009). 
Phookun et al. (1993) separated an outer 
H{\sc i} ring structure from the emission of the galactic disk. They suggested that this ring structure is made of
backfalling gas.  The Nobeyama CO(1--0) distribution is consistent with the IRAM 30m CO(2--1) distribution presented in Schruba et al. (2011).
NGC~4254 is less molecular-gas rich than NGC~4321. The star formation rate is approximately proportional
to the molecular gas density down to our detection limit of $5$~M$_{\odot}$pc$^{-2}$.
The scatter increases from $0.1$~dex at high surface densities to 0.5~dex at our detection limit.
The star formation rate is proportional to the total gas surface density where the molecular gas dominates
(down to $\sim 20$~M$_{\odot}$pc$^{-2}$). In the H{\sc i}-dominated region below this value the star formation efficiency $SFE_{\rm tot}$
decreases significantly and the correlation between the gas surface density and the star formation rate becomes poor.
There are regions in the northern outer H{\sc i} ring structure where $SFE_{\rm tot} \sim SFE_{\rm HI}$ is significantly smaller (by a factor of $2$--$3$)
 than within the disk.

NGC~4321 (Fig.~\ref{fig:plot_n4321}): this is the only galaxy in the sample which resembles closely a ``normal'' field spiral galaxy.
The H{\sc i} distribution is symmetric and extends beyond the optical radius. Its radio continuum properties
including polarization are also normal (Vollmer et al. 2007). The Nobeyama CO(1--0) distribution is 
consistent with the IRAM 30m CO(2--1) distribution presented in Schruba et al. (2011). 
The galaxy is molecular-gas rich (Wong \& Blitz 2002). The molecular gas fraction ($\Sigma_{{\rm H}_{2}}/\Sigma_{\rm HI}$) 
is unity at about $0.7 \times R_{25}$.
The star formation rate is proportional to the molecular gas surface density with a scatter of $0.2$--$0.3$~dex.
This proportionality seems also to hold for the total gas surface density, although with our CO detection limit of
$\sim 4$~M$_{\odot}$pc$^{-2}$ we miss CO emission below a total gas surface densities of $\sim 10$~M$_{\odot}$pc$^{-2}$. 

NGC~4330 (Fig.~\ref{fig:plot_n4330}): the H{\sc i} distribution is truncated within the optical disk to the northeast and has a low surface density
tail to the southwest (Chung et al. 2009) without a stellar counterpart. No CO is detected in this gas tail (Vollmer et al. 2011).
A UV tail is present in the southwest offset from the gas tail
(Abramson et al. 2011). According to the dynamical model of Vollmer et al. (2011) the galaxy is moving to the north.
This is supported by the location of a radio-deficit region (Murphy et al. 2009). 
The molecular fraction is highest at this side of the galaxy. $SFE_{\rm tot}$ decreases 
by a factor of $\sim 2$ along the direction of the galaxy's minor axis from the northwest to the southeast. 
This gradient is also observed in the gas tail.
The star formation rate is proportional to the molecular gas surface density down to a surface density of
1~M$_{\odot}$pc$^{-2}$ with a small scatter of 0.1~dex. Because of the high inclination of the galactic disk, the correlation between the
atomic/total gas surface density and star formation is difficult to interpret because of projection effects.

NGC~4402 (Fig.~\ref{fig:plot_n4402}): the H{\sc i} disk is truncated well inside the stellar disk with a small H{\sc i} tail to the northwest. 
This galaxy also shows a compressed radio continuum halo (Crowl et al. 2005) and a
polarized radio continuum ridge to the south. The galaxy thus undergoes ram pressure stripping from the south
(Crowl et al. 2005). The gas disk is mostly dominated by the molecular phase and star formation is approximately proportional to the
molecular and total gas surface density with a scatter of $0.2$--$0.3$~dex.

NGC~4419 (Fig.~\ref{fig:plot_n4419}): as in NGC~4402 the H{\sc i} disk is strongly truncated (Chung et al. 2009). 
Within the disk there is more gas to the northwest of the 
galaxy center than in the southeast as already pointed out by Kenney et al. (1990). This gas is dominated by the molecular phase. 
Because of the high infrared surface brightness of the nucleus which lead to PSF artifacts, we had to clip the image of the star formation rate
at $\dot{\Sigma}_{*} < 5 \times 10^{-3}$~M$_{\odot}$kpc$^{-2}$yr$^{-1}$.
The galaxy nucleus has a very high gas surface density and star formation rate. 
The star formation rate is proportional to the molecular gas surface density with a scatter of 0.5~dex.

NGC~4438 (Fig.~\ref{fig:plot_n4438}) had a strong gravitational interaction $\sim 100$~Myr ago (Vollmer et al. 2005, Kenney et al. 2008). 
It now undergoes strong active ram pressure stripping (Vollmer et al. 2005) leading to double line profiles in the strong western extraplanar 
CO emission. The gas content of NGC~4438 is dominated by the molecular phase. The star formation rate of the
extraplanar region based on the H$\alpha$ emission does not correlate with the total gas distribution (Fig.~\ref{fig:plot_n4438_ha}).
The star formation contours follow the gas contours only at the northern tip of the extraplanar region.
The reason for this behavior is that most of the H$\alpha$ emission in the extraplanar region stems from shock-ionized gas and not from H{\sc ii}
regions (Kenney \& Yale 2002, Vollmer et al. 2009). We therefore rely on the star formation map based on UV emission.
At molecular gas surface densities $> 15$~M$_{\odot}$pc$^{-2}$
the star formation rate is proportional to the molecular gas surface density. At smaller molecular gas surface densities
the correlation steepens. The star formation efficiency with respect to the molecular gas $SFE_{\rm mol}=\dot{\Sigma}_{*}/\Sigma_{\rm mol}$ 
decreases by about a factor of $3$ between the disk and the extraplanar regions. 

NGC~4501 (Fig.~\ref{fig:plot_n4501}): the H{\sc i} distribution is truncated at about the optical radius $R_{25}$ with a sharp edge to the southeast and
more widespread H{\sc i} emission distribution at the opposite side (Chung et al. 2009). The polarized radio continuum emission
shows a ridge to the southeast indicating that the ram pressure is acting on the ISM from this side (Vollmer et al. 2008a).
This galaxy is molecular-gas rich (Wong \& Blitz 2002) and has a central H{\sc i} depression. Whereas the molecular fraction decreases along the minor
axis to the northeast, it stays at a high level at the southwestern edge. $SFE_{\rm tot}$ is uniform across
the entire galactic disk except in the southeast where it is a factor of $2$--$3$ lower than in the disk. 
Whereas the total gas surface density is quite high
($\sim 20$~M$_{\odot}$pc$^{-2}$) in this region, the star formation rate is low ($2 \times 10^{-3}$~M$_{\odot}$kpc$^{-2}$yr$^{-1}$).
In the inner part of the galactic disk, where the molecular gas surface density is higher than 40~M$_{\odot}$pc$^{-2}$, the star formation 
rate is almost constant with increasing molecular gas surface density. At lower gas surface densities the correlation is 
linear with a large scatter (0.8~dex). An important part of this scatter is due to the low signal-to-noise ratio of the CO data.
The star formation rate appears to be approximately proportional to the total gas surface density with an extremely large scatter of 0.7~dex. 

NGC~4522 (Fig.~\ref{fig:plot_n4522}): has a strongly truncated H{\sc i} distribution with high column density extraplanar gas to the northwest (Kenney et al. 2004).
The H$\alpha$ emission distribution is truncated at the same place in the disk as the H{\sc i} distribution, 
whereas the FUV distribution extends beyond the H{\sc i} truncation radius. This is because 
NGC~4522 was stripped very recently (several $10$~Myr; Crowl \& Kenney 2006, Vollmer et al. 2006) and the timescale of gas stripping due to ram pressure 
is shorter than the timescale of FUV emission ($\sim 100$~Myr).  
Thus beyond the H{\sc i} truncation radius, there is significant FUV emission from stars younger than $\sim 100$~Myr, although there is no 
ongoing star formation (within the last $\sim 10$~Myr) here. However, it is still possible to derive a meaningful star formation rate from either FUV or H$\alpha$ 
in regions with HI emission, because FUV emission from ongoing star formation dominates the total FUV in these regions.
The CO emission within the disk is more extended to 
the northeast, whereas only the southwestern extraplanar H{\sc i} emission blob shows a detectable molecular component.
We also note that the molecular fraction along the minor axis is higher toward the southeast. Similar to NGC~4330, $SFE_{\rm tot}$ 
of the extraplanar gas is about $3$--$4$ times lower than that of the gas in the disk. 
The star formation rate is proportional to the molecular gas surface density down to $1$~M$_{\odot}$pc$^{-2}$ with a small scatter ($\leq 0.3$~dex). 
Because of the high inclination of the galactic disk, the correlation between the
atomic/total gas surface density and star formation is difficult to interpret because of projection effects.

NGC~4548 (Fig.~\ref{fig:plot_n4548}): is an anemic galaxy with a low overall H{\sc i} gas surface density, a central H{\sc i} hole (Cayatte et al. 1994), and
a low star formation rate (Koopmann \& Kenney 2004). Moreover, the galaxy has a prominent bar. 
Moderate star formation is detected in the center and in two spiral arms beginning at the end of the bar.
The highest gas surface densities are found in the center, the bar, and two spiral arms beginning at the end of the bar.
The gas distribution is truncated at the optical radius.
The poor signal-to-noise of the CO data at these low molecular gas surface densities does not permit to draw firm conclusions on the
spatial distribution of the star formation efficiency. Despite a large scatter ($0.6$~dex), the star formation is proportional
to the molecular gas surface density except in the bar region where the star formation rate is low compared to the molecular
gas surface density. As in NGC~4501 the correlation between star formation and total gas surface density is poor.

NGC~4569 (Fig.~\ref{fig:plot_n4569}): has a strongly truncated gas disk with a one-sided arm structure to the west (Vollmer et al. 2004a, Chung et al. 2009).
This arm structure consists of resettling ISM which has been pushed to larger galactic radii by a ram pressure
stripping event $\sim 300$~Myr ago (Vollmer et al. 2004a, Boselli et al. 2006).
The Nobeyama CO(1--0) distribution is consistent with the IRAM 30m CO(2--1) distribution presented in Schruba et al. (2011).
CO is detected at the northern tip of this arm. Within the disk the molecular fraction is higher on the southern side. There,
$SFE_{\rm tot}$ is $\sim 2$ times lower than north of the galaxy center. The star formation rate in the western arm is
not significantly different from that of the disk. The star formation rate is proportional to the molecular gas surface density and
the total gas surface density with a scatter of $0.6$~dex, because the ISM is dominated by the molecular component.

NGC~4579 (Fig.~\ref{fig:plot_n4579}): is the second anemic galaxy in our sample. As in NGC~4548 it has a prominent bar and a low overall low gas surface density and
star formation rate. The highest gas surface densities are found in two spiral arm departing from the end of the bar.
In the smoothed image this leads to a quasi ring structure of the gas distribution. The gas distribution is truncated at the optical radius.
Despite the low total gas surface density
the molecular fraction is as high as in a ``normal'' galaxy like NGC~4321. The star formation rate is proportional to the 
molecular gas surface density with a scatter of $0.3$~dex. 
Since the molecular fraction is high in the entire disk, the correlation between star formation
and the total gas surface density is only slightly steeper than that of the molecular gas surface density.
The gap around $\Sigma_{\rm g}=4$~M$_{\odot}$pc$^{-2}$ indicates that we are missing CO emission below our limit of 
$\Sigma_{\rm H_{2}}=4$~M$_{\odot}$pc$^{-2}$.

NGC~4654 (Fig.~\ref{fig:plot_n4654}): has a very extended H{\sc i} distribution with a southeastern low surface density tail (Phookun \& Mundy 1995).
The star formation distribution also shows two small tails in this direction. 
Dynamical modelling suggests that this galaxy had a rapid tidal interaction in the past and is now undergoing low ram pressure stripping (Vollmer 2003).
The distribution of the molecular gas is asymmetric
with more CO to the northwestern sharp edge of the H{\sc i} distribution. $SFE_{\rm tot}$ is high in the galaxy
center and the northwestern gas surface density maximum. It is particularly low in the southeast. The star formation is
approximately proportional to the molecular gas surface density with a large scatter ($0.7$~dex) due to the poor signal-to-noise ratio
of the CO data.

\section{Discussion \label{sec:discussion}}

The recent THINGS (Walter et al. 2008) and HERACLES (Leroy et al. 2009) surveys of local spiral galaxies provide an excellent 
comparison to our Virgo cluster sample.

\subsection{Gas content and molecular fraction \label{sec:rmol}}

\begin{figure}
  \centering
   \resizebox{\hsize}{!}{\includegraphics{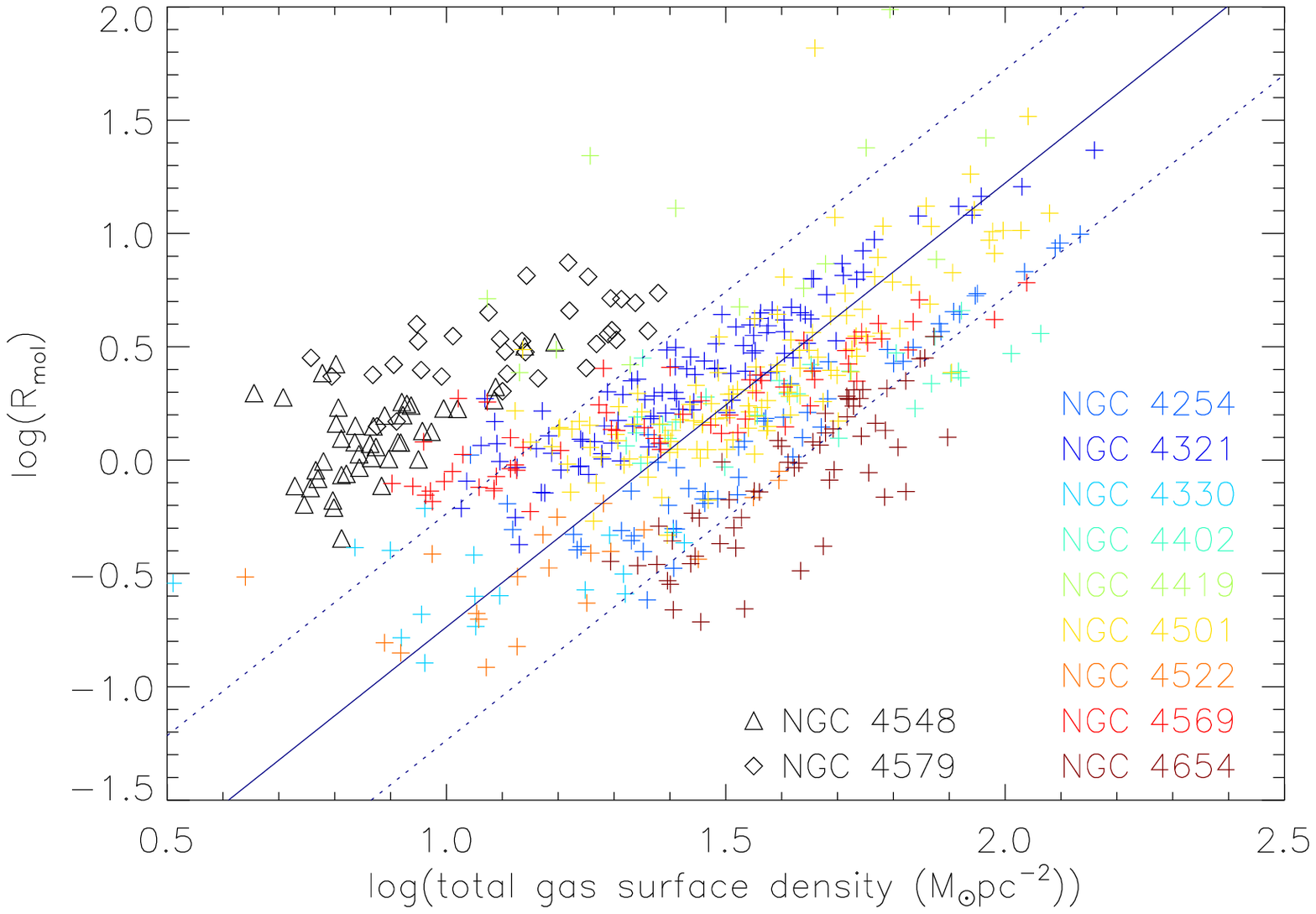}}
   \resizebox{\hsize}{!}{\includegraphics{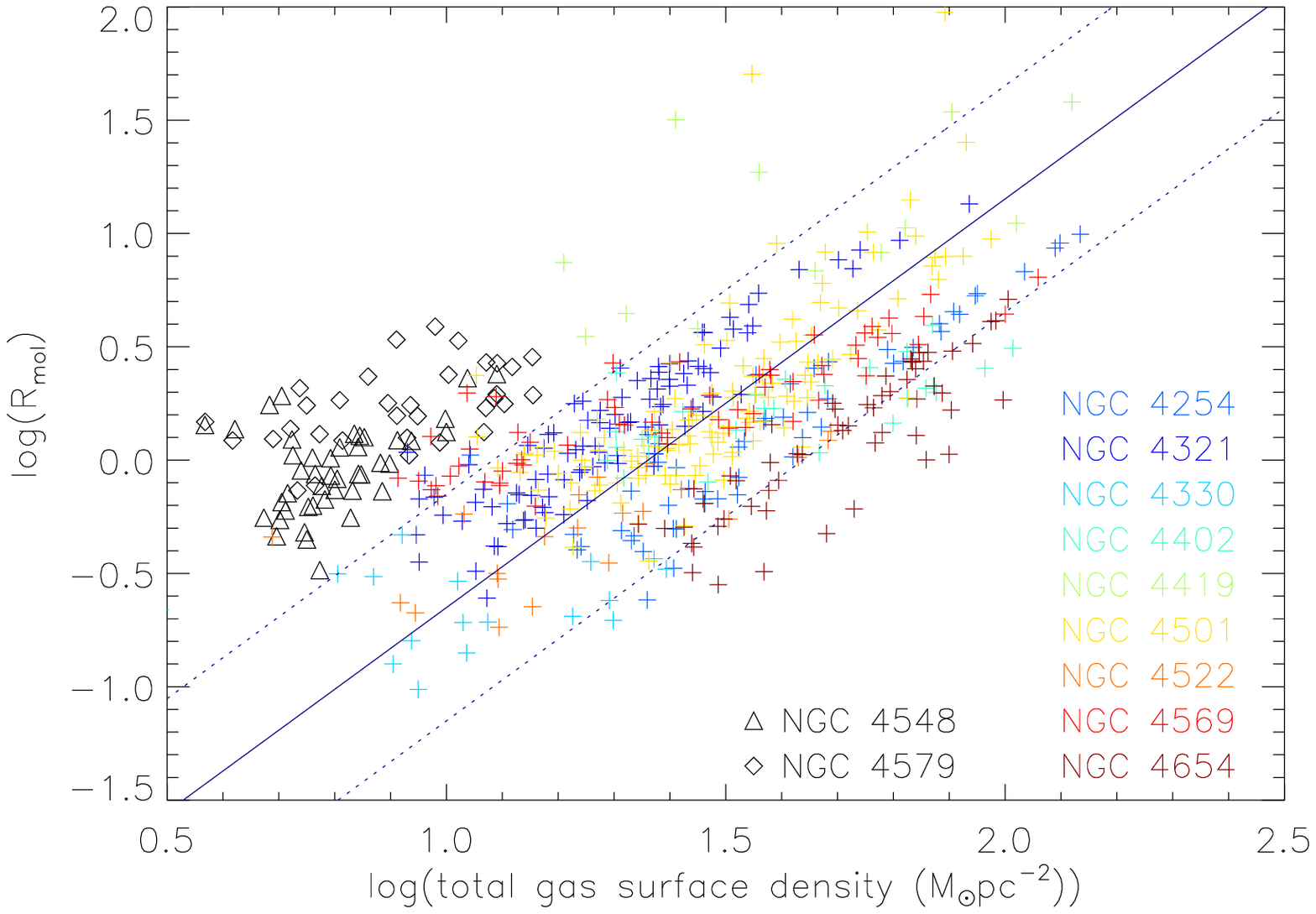}}
  \caption{Molecular fraction $\Sigma_{\rm H_{2}}/\Sigma_{\rm HI}$ as a function of the total gas surface density 
    $\Sigma_{\rm g}=\Sigma_{\rm H_{2}} + \Sigma_{\rm HI}$ for all sample galaxies except NGC~4438.
    Upper panel: data with a constant CO-to-H$_{2}$ conversion factor. Lower panel: with SFR/L(CO)-normalized molecular gas surface densities (see text).
    Symbols of different colors represent different galaxies. Black symbols represent the two anemic galaxies NGC~4548 (triangles) 
    and NGC~4579 (diamonds). The solid lines represent a linear regression to the data of all galaxies except NGC~4548 and NGC~4579 using
    a Bayesian approach to linear regression with errors in both variables (Kelly 2007). To guide the eye, the dotted lines
    represent offsets of a factor of two.
  \label{fig:rmol_graph}}
\end{figure}

The molecular and atomic gas content of the inner disks of our galaxy sample is not significantly different from 
that of local galaxies (Schruba et al. 2011). There are galaxies with high H$_{2}$ surface densities and H{\sc i} depressions in the center:
NGC~4321, NGC~4419, NGC~4501. This kind of gas morphology in the inner galactic disk is also present in
the field, e.g., NGC~628, NGC~3351, NGC~5457. The central gas disks of NGC~4254, NGC~4402, NGC~4569, and NGC~4654 have
high molecular gas surface densities with H{\sc i} gas surface densities close to the typical value of $5$-$10$~M$_{\odot}$pc$^{-2}$
(Leroy et al. 2008). Comparable nearby field galaxies are NGC~3521, NGC~5055, and NGC~7331.
The molecular and atomic gas surface densities of the inner disks of NGC~4330 and NGC~4522 are intermediate\footnote{The surface densities are
not corrected for inclination.} 
($\sim 10$-$30$~M$_{\odot}$pc$^{-2}$) with a molecular fraction of $\Sigma_{\rm H_{2}}/\Sigma_{\rm HI} \sim 1$. Since these two galaxies are
edge-on, projection effects decrease the molecular fraction, because the inner disk measures are blended with the H{\sc i}-rich
outer parts.
For comparison, the nearby spiral galaxies NGC~2403, NGC~925, and NGC~3198 also show a molecular fraction around unity in the inner disk.
There are two anemic galaxies with smooth spiral arms and characteristics between S0s and normal spirals in our sample: NGC~4548 and NGC~4579. 
Both have a low gas surface density everywhere together with a high molecular fraction 
$R_{\rm mol} = \Sigma_{\rm H_{2}}/\Sigma_{\rm HI} \geq 1$ (Fig.~\ref{fig:rmol_graph}). 

The global (without NGC~4548 and NGC~4579) unnormalized and normalized $R_{\rm mol}$-$\Sigma_{\rm g}$ relations have a Spearman rank coefficient of $0.7$.
As for the star formation rate (Sect.~\ref{sec:results}), we performed a linear regression to the initial and normalized data assuming
a constant error of $0.3$~dex for the molecular gas surface density and a negligible error for the H{\sc i} surface density.
The global (without NGC~4548 and NGC~4579) unnormalized and normalized $R_{\rm mol}$-$\Sigma_{\rm g}$ relations have slopes of 
$2.0 \pm 0.2$ and $1.8 \pm 0.2$, respectively.
Within most individual galaxies the relationship between $R_{\rm mol}$ and $\Sigma_{\rm g}$ is between linear and quadratic 
with a scatter of about 0.5~dex. But there is a much larger spread among galaxies. Although these values are not corrected for galaxy inclination, we note that the 
galaxies with extreme values of $R_{\rm mol}$ are all moderately inclined, so the spread of $R_{\rm mol}$ values with respect to gas density are intrinsic. 
Of particular interest are the anemic galaxies NGC~4548 and NGC~4579. Whereas the absolute molecular fractions of NGC~4548 and NGC~4579 are 
similar to those of the other galaxies, we observe a significant offset from the $R_{\rm mol}$-$\Sigma_{\rm g}$ relation of the ten other sample galaxies. 
This is contrary to the
claim of Fumagalli et al. (2009) that the molecular fraction depends primarily on the total gas surface density (see also Gardan et al. 2007).
A field counterpart of the two anemic cluster spirals is NGC~2841. Schruba et al. (2011) pointed out that early type galaxies,
as NGC~2841, also display high molecular fractions at low total gas column densities. Indeed, NGC~2841, NGC~4548, and NGC~4579 are all Sb galaxies.
Our results show that the molecular fraction depends on more than the total gas surface density. We suggest that the observed
enhanced molecular fraction with respect to the total gas surface density is caused by an increased pressure due to the stellar surface density
(Elmegreen 1993, Wong \& Blitz 2002).

Among the Virgo spiral galaxies of our sample NGC~4419, NGC~4402, NGC~4569, NGC~4330, and NGC~4522 show gas disks which are 
strongly truncated inside the optical disk. However, their inner gas disks have the same gas properties as the inner regions of normal spirals
(see above). Since for all Virgo spirals of our sample we can find nearby galaxy counterparts for inner disk H{\sc i} and CO distributions, 
we conclude that gas truncation does not necessarily lead to a major change in the gas surface density distribution of the inner disk of Virgo 
spiral galaxies (see also Fumagalli et al. 2009 for the radial profiles). On the other hand, molecular gas depletion occurs in field and cluster spirals.
Four field and five cluster galaxies of the Fumgalli et al. (2009) sample (see their Table 1) are gas (H{\sc i} and CO) deficient.

\begin{figure*}
  \centering
  \includegraphics[width=14cm]{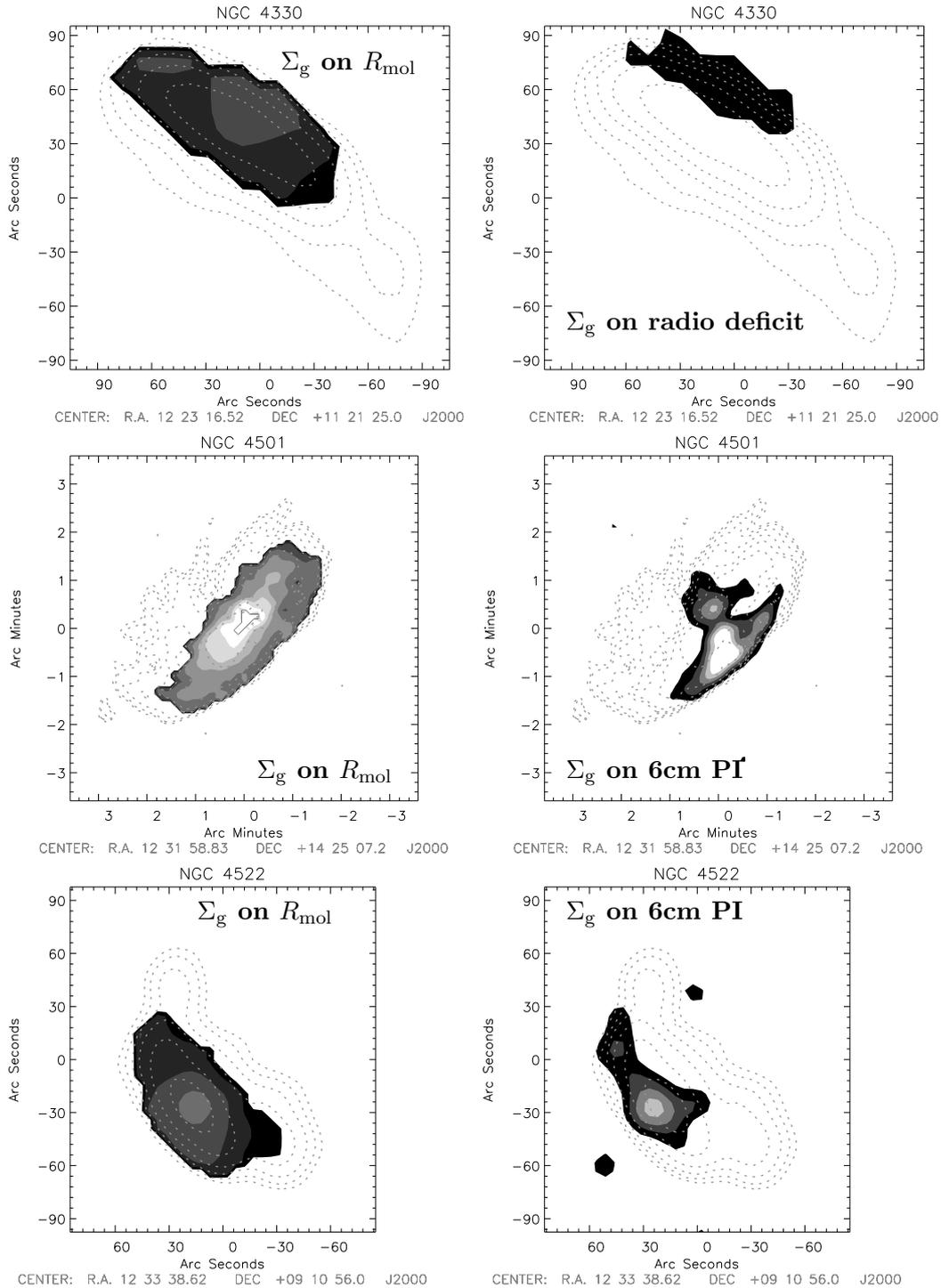}
  \put(-280,520){\bf \large $\Sigma_{\rm g}$ on $R_{\rm mol}$}
  \put(-290,220){\bf \large $\Sigma_{\rm g}$ on $R_{\rm mol}$}
  \put(-315,160){\bf \large $\Sigma_{\rm g}$ on $R_{\rm mol}$}
  \put(-160,410){\bf \large $\Sigma_{\rm g}$ on radio deficit}
  \put(-160,220){\bf \large $\Sigma_{\rm g}$ on 6cm PI}
  \put(-160,160){\bf \large $\Sigma_{\rm g}$ on 6cm PI}
  \caption{The three galaxies which show an enhanced molecular fraction at the windward side. Left panels: 
    Total gas surface density contours on the molecular gas fraction $\Sigma_{\rm H_{2}}/\Sigma_{\rm HI}$.
    Greyscale levels are $(1,2,4,8,16,32) \times 0.1$. Right panels: Total gas surface density contours on 
    the radio deficit region in greyscale (NGC~4330; from Murphy et al. 2009) or the 6cm polarized radio continuum
    emission distribution (NGC~4501; from Vollmer et al. 2007; NGC~4522; from Vollmer et al. 2004b).
  \label{fig:rmol_pi}}
\end{figure*}

Can we identify a change in the molecular fraction due to ram pressure? Five galaxies are affected by an active ram pressure
wind: NGC~4330 (Abramson et al. 2011), NGC~4402 (Crowl et al. 2005), NGC~4438 (Vollmer et al. 2005b),
NGC~4501 (Vollmer et al. 2008a), NGC~4522 (Kenney et al. 2004, Vollmer et al. 2006), and NGC~4654 (Vollmer 2003). 
In addition, NGC~4569 had its peak ram pressure $\sim 300$~Myr ago (Vollmer 2009) and is still affected by active ram pressure 
(Murphy et al. 2009, We\.{z}gowiec et al. 2011).
All these galaxies but NGC~4654 have strongly truncated gas disks, and CO is easily detected at the gas truncation radii, which are well inside 
the optical radii. But NGC~4654's gas disk is truncated at about the optical radius, where CO emission is generally weak, and while strong CO is 
detected in the inner disk, the CO observations of NGC~4654 did not have enough sensitivity to detect any faint CO out this far in the disk.
Only NGC~4330, NGC~4501, and NGC~4522 show a $1.5$--$2$ times enhanced molecular fraction on the windward side with respect to the leeward 
side (Fig.~\ref{fig:rmol_pi}). 
In NGC~4330 and NGC~4522, projection effects make the interpretation of the high $R_{\rm mol}$ along the minor axis difficult, 
because the atomic gas surface density within the galactic disk at the leeward 
side is polluted by stripped extraplanar gas. In NGC~4330 there are two regions of high $R_{\rm mol}$: (i) to the north along the minor axis
and (ii) to the northeast of the galaxy center at the end of the major axis. The asymmetry with respect to the minor axis 
indicates that the latter region displays an enhanced molecular fraction. Therefore,
the only unambiguous candidate regions for a 
ram-pressure enhanced molecular fraction are the northeastern upturn region in NGC~4330 (Fig.~\ref{fig:plot_n4330})
and the southwestern edge of NGC~4501's gas disk (Fig.~\ref{fig:plot_n4501}). The interpretation of this effect is that ram pressure compresses
the gas which becomes denser leading to a higher molecular fraction. Another case of a face-on viewed galaxy undergoing active ram
pressure stripping is needed to confirm our suggestion.

\subsection{Star formation efficiency}

The ratio between star formation rate and molecular gas surface density is about constant for all but two Virgo cluster galaxies
as shown by Fumagalli et al. (2008) for the radial profiles.
The exceptions are NGC~4501, where the star formation rate increases much more slowly toward the galaxy center than the molecular
gas surface density, and NGC~4438, where the star formation efficiency $SFE_{\rm tot} = SFE_{\rm mol}$ of the extraplanar gas is $\sim 3$ times lower
than that of the galactic disk.
The scatter of the molecular gas consumption rate ($\sim 0.3$~dex) is comparable to that derived from the
nearby field spiral sample $0.24$~dex (Bigiel et al. 2008, Leroy et al. 2008, Schruba et al. 2011). A significant part of this
may be explained by a galaxy-to-galaxy variation of the CO-to-H$_2$ conversion factor (Schruba et al. 2011).
We thus conclude that in most cases the cluster environment does not affect the resolved 2D star formation efficiency with respect to the
molecular gas, confirming the results of Fumagalli et al. (2008) based on radial profiles.
However, in an extreme case as NGC~4438 extraplanar gas can show a significantly decreased $SFE_{\rm mol}$ (Fig.~\ref{fig:sfr}). 
For galaxies with symmetric gas distributions the correlation between the star formation rate and the total gas surface density is as tight as that
between the star formation rate and the molecular gas surface density (Table~\ref{tab:corr}).
The rank correlation between the star formation rate and the total gas surface density is smaller than that
between the SFR and the molecular gas surface density for galaxies with significant extraplanar H{\sc i} emission (NGC~4330, NGC~4522),
because of a decreased star formation rate in the extraplanar gas. 
The opposite trend is observed for NGC~4654, although this may be an artifact caused the relatively poor quality of the CO data.
The correlation between the star formation rate and the molecular gas surface density using a constant CO-to-H$_{2}$ conversion factor
is as tight as the correlation between the star formation rate and the total gas surface density.
The correlation of the
star formation rate and the SFR/L(CO)-normalized molecular gas surface density is tighter than the SFR-$\Sigma_{\rm g}$ correlation (Table~\ref{tab:corr}).
Our findings are in agreement with those of Schruba et al. (2011) and in contrast to the findings of 
Fumagalli et al. (2008). We suspect that the latter authors based their claim on CO emission at radii larger
than $0.5 \times R_{25}$ where the detection is uncertain.

\begin{table}
      \caption{Spearman rank correlation between the star formation rate and the molecular/total gas surface density.}
         \label{tab:corr}
      \[
         \begin{array}{lcc}
           \hline
           \noalign{\smallskip}
           {\rm galaxy\ name\ \ \ \ \ \ } & {\rm molecular} & {\rm total} \\
	   \noalign{\smallskip}
	   \hline
	   \noalign{\smallskip}
	   {\rm NGC~4254} & 0.90 & 0.94 \\            
	   {\rm NGC~4321} & 0.96 & 0.97 \\           
	   {\rm NGC~4330} & 0.98 & 0.90 \\	  
	   {\rm NGC~4402} & 0.94 & 0.92 \\	  
	   {\rm NGC~4419} & 0.81 & 0.74 \\	   
	   {\rm NGC~4501} & 0.82 & 0.88 \\	  
	   {\rm NGC~4522} & 0.97 & 0.89 \\	  
	   {\rm NGC~4548} & 0.48 & 0.37 \\
	   {\rm NGC~4569} & 0.85 & 0.93 \\	   
	   {\rm NGC~4579} & 0.88 & 0.92 \\	   
	   {\rm NGC~4654} & 0.67 & 0.89 \\
	   \noalign{\smallskip}
	   \hline
	   \noalign{\smallskip}
	   {\rm all\ galaxies\ (constant\ CO\ to\ H_{2}\ factor)} & 0.86 & 0.84 \\
	   {\rm all\ galaxies\ (normalized)} & 0.92 & 0.83 \\
	   \noalign{\smallskip}
           \hline
        \end{array}
      \]
\end{table}

In normal (non-anemic) nearby field spiral galaxies the transition between a molecular-gas-dominated and atomic-gas-dominated ISM 
occurs at $14 \pm 6$~M$_{\odot}$pc$^{-2}$ or $0.43 \pm 0.18 \times R_{25}$ (Leroy et al. 2008)\footnote{The quoted gas surface density is 
corrected for inclination, whereas our gas surface densities
are not corrected for inclination.}. For lower gas surface density,
where the ISM is mostly atomic, the correlation between star formation and gas surface density is weak. 
Regions of low gas surface densities ($\leq 5$~M$_{\odot}$pc$^{-2}$) are typically located outside the optical
radius ($R_{25}$). In these regions $SFR_{\rm tot} \sim SFR_{\rm HI}$ and $SFE_{\rm tot} \sim SFE_{\rm HI}$ are 
low ($SFR_{\rm tot} \leq 10^{-4}$~M$_{\odot}$kpc$^{-2}$yr$^{-1}$ and $SFE_{\rm tot} \leq 10^{-10}$~yr$^{-1}$;
Bigiel et al. 2010). Only NGC~4254 and NGC~4654, which have gas disks extending beyond the optical radius, 
show this behavior. In all other galaxies, the star formation rate in regions inside the optical disk
with low total gas surface densities show relatively high star formation efficiencies ($SFE_{\rm tot}=2.5$-$5 \times 10^{-10}$~yr$^{-1}$).

Two galaxies in our sample are of particular interest: NGC~4548 and NGC~4579. Both are classified as anemic, i.e. their total gas surface density
and star formation rate is low all over the galactic disk (Cayatte et al. 1994, Koopmann \& Kenney 2004).
Here we show that the reason for the anemic starforming disks is that they have a low total gas surface density.
These galaxies show molecular fractions of $R_{\rm mol} \sim 1$ (Sect.~\ref{sec:rmol}) despite their low total gas surface densities.
Whereas $SFE_{\rm mol}$ is normal, their total star formation efficiency is relatively high ($SFE_{\rm tot} \sim 2$-$5 \times 10^{-10}$~yr$^{-1}$),
because of the low H{\sc i} surface density.
Anemic galaxies are also found in the field (Elmegreen et al. 2002): nearly all Sa galaxies with extended star formation over the whole galactic disk, 
even those in the field, were classified as anemic by van den Bergh (1976).
The closest nearby galaxy counterpart is the Sb galaxy NGC~2841 which has a low gas surface density over the whole disk.
Its total star formation efficiency with respect to the total gas is about $3 \times 10^{-10}$~yr$^{-1}$. 

The difference between NGC~2841 and the two anemic Virgo galaxies is that NGC~2841's gas surface density remains constant beyond the 
optical radius, whereas the gas surface density of the two Virgo anemic galaxies clearly declines between $0.5 \times R_{25}$ and $R_{25}$
(Chung et al. 2009, Schruba et al. 2011). Such a decline is also 
observed in the outer disk ($R > 0.5 \times R_{25}$) of the nearby spiral galaxy NGC~3627, 
which is part of a triplet and has a ``classical'' high column density gas distribution in the inner disk.
Despite this decline the star formation efficiency $SFE_{\rm tot}$ stays approximately constant in NGC~3627 (Schruba et al. 2011), a behavior
which is also observed in the Virgo galaxies NGC~4548 and NGC~4579.
Since we find the same behavior of the star formation efficiency $SFE_{\rm tot}$ at low gas surface densities inside the
optical disk in the field and cluster environment, we conclude that the relatively high star formation efficiencies of NGC~4548 and
NGC~4579 are not due to the cluster environment. The most probable explanation for the high star formation
efficiency $SFE_{\rm tot}$ is a high metallicity and an enhanced gas volume density or pressure due to a higher stellar surface density via the gravitational potential 
(Vollmer \& Leroy 2011) leading to a smaller free-fall time. The presence of a strong bar in NGC~3627, NGC~4548, and NGC~4579 might be related to their
decreasing gas surface density inside the optical disk.
To explain the observed star formation efficiencies in our Virgo cluster sample, we do not need to invoke
an additional external thermal pressure (not ram pressure) from intracluster medium as suggested by Nakanishi et al. (2006).

\begin{figure*}
  \centering
  \includegraphics[width=12cm]{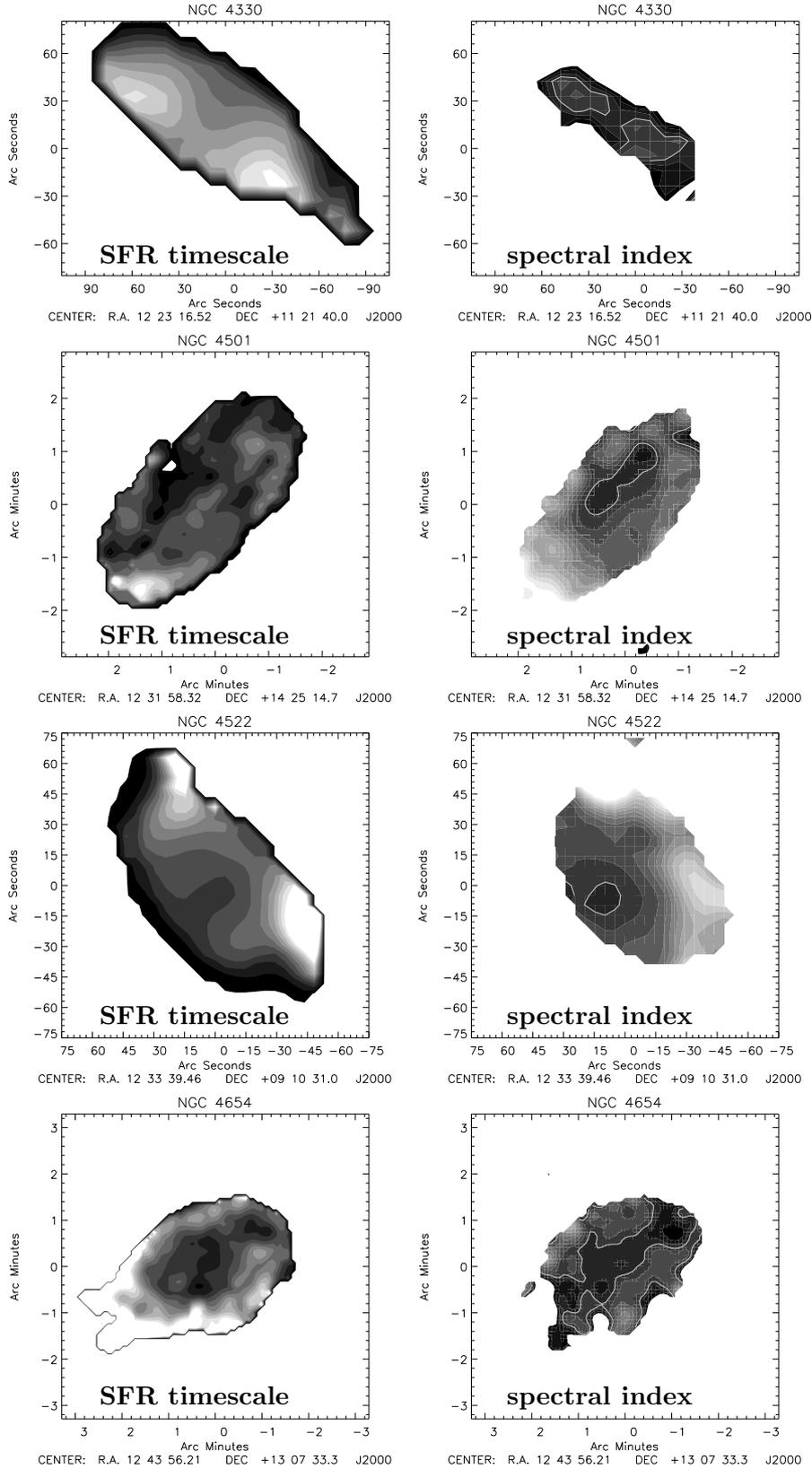}
   \put(-300,510){\bf \large SFR timescale}
   \put(-300,350){\bf \large SFR timescale}
   \put(-300,190){\bf \large SFR timescale}
   \put(-300,30){\bf \large SFR timescale}
   \put(-130,510){\bf \large spectral index}
   \put(-130,350){\bf \large spectral index}
   \put(-130,190){\bf \large spectral index}
   \put(-130,30){\bf \large  spectral index}
  \caption{Four galaxies which show asymmetric regions of decreased star formation efficiency. Left panels: 
    star formation timescale $t_{*}=\Sigma_{\rm g}/\dot{\Sigma}_{*}$.
    Greyscale levels are $(1,10,20,30,40,50,60,70,80,90,100)$~10~Myr. Right panels: spectral index distribution between
    $4.8$~GHz and $1.4$~GHz (NGC~4501 and NGC~4654 from  Vollmer et al. 2010; NGC~4522 from Vollmer et al. 2004b).
    The greyscale levels are from $0.5$ to $2.0$ in steps of $0.1$. The additional contour corresponds to a spectral index of $0.8$.
  \label{fig:sft_si}}
\end{figure*}

\begin{figure}
  \centering
  \resizebox{\hsize}{!}{\includegraphics{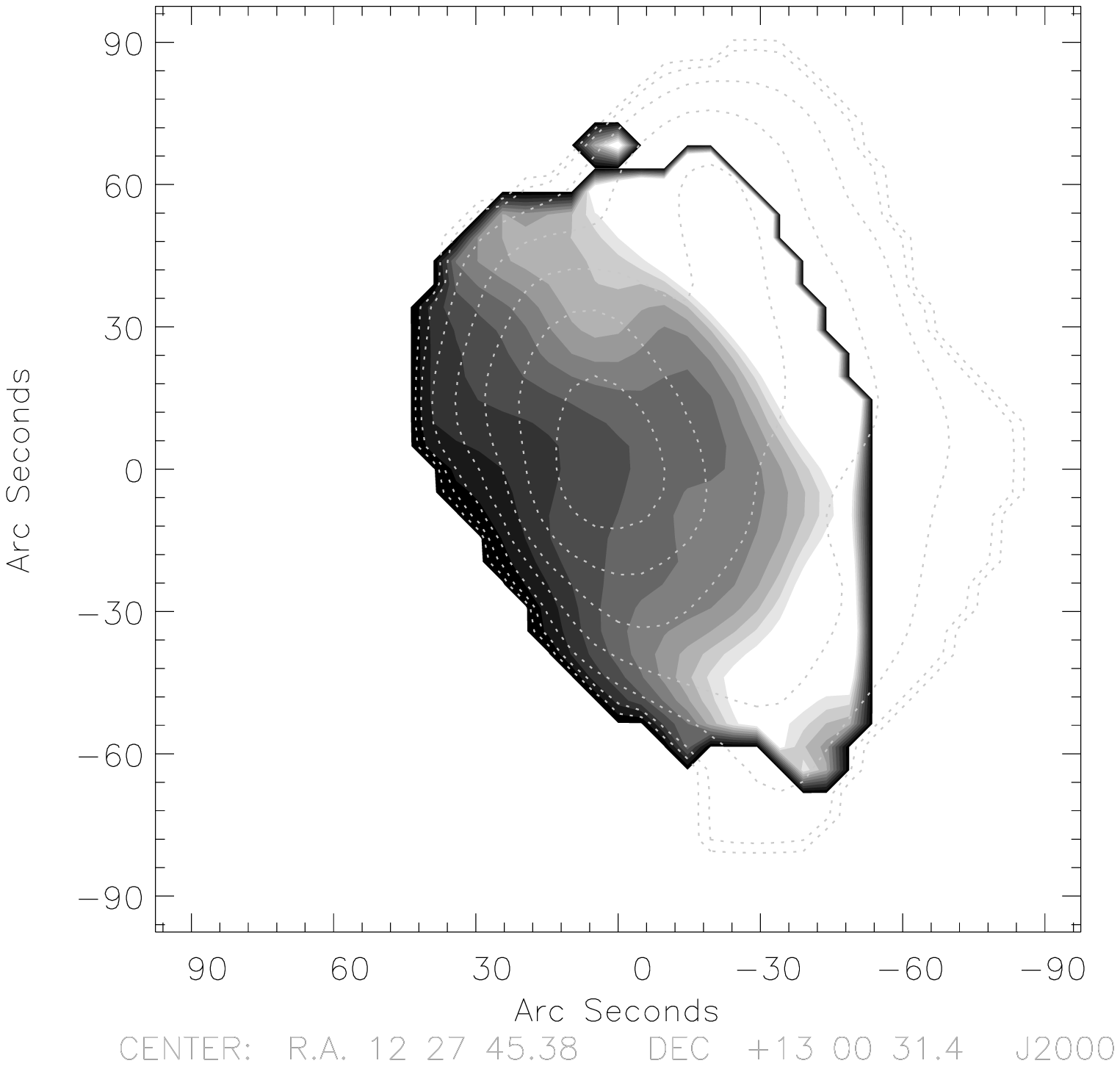}}
  \put(-200,50){\bf \large $\Sigma_{\rm g}$ on SFR timescale}
  \caption{NGC~4438, which shows asymmetric regions of decreased star formation efficiency. 
    total gas surface density contours on the star formation timescale $t_{*}=\Sigma_{\rm g}/\dot{\Sigma}_{*}$.
    Greyscale levels are $(1,10,20,30,40,50,60,70,80,90,100)$~10~Myr. 
  \label{fig:n4438_sft_si}}
\end{figure}

We observe an increase of the star formation efficiency with respect to the total gas $SFE_{\rm tot}$ associated with
the regions of high molecular fractions (see Sect.~\ref{sec:rmol}) in NGC~4330 (Fig.~\ref{fig:plot_n4330}) and NGC~4522 (Fig.~\ref{fig:plot_n4522}),
but not in NGC~4501 (Fig.~\ref{fig:plot_n4501}). 
There is no significant increase of the star formation efficiency $SFE_{\rm tot}$ on the windward side of ram pressure-stripped
galaxies NGC~4402, NGC~4569, and NGC~4654. 
The main difference between the distributions of the star formation efficiency in nearby field and Virgo cluster spirals
are asymmetric regions of very low star formation efficiencies with respect to the total gas ($SFE_{\rm tot} < 10^{-10}$~yr$^{-1}$) in NGC~4254, NGC~4330, 
NGC~4438, NGC~4501, NGC~4522, and NGC~4654. 
These regions show a $2$--$4$ times lower star formation efficiency $SFE_{\rm tot}$ than regions at the
opposite side of the galactic disk and are clearly associated with extraplanar gas in the edge-on spiral galaxies
NGC~4330, NGC~4438, and NGC~4522. 
Phookun et al. (1993) also claimed that the northern extended gas disk of NGC~4254 is backfalling,
extraplanar gas. Since this gas is located beyond the optical radius of NGC~4254, low star formation efficiencies are expected in these
regions. 

The southeastern region of low star formation efficiency $SFE_{\rm tot}$ in NGC~4501 is located inside the
optical radius and coincides with regions of steep spectral indices ($< -1$) between $1.4$ and $4.8$~GHz (Vollmer et al. 2010). 
A steepening of the radio spectral index is also found in the extraplanar gas of NGC~4522 (Vollmer et al. 2004b; Fig.~\ref{fig:sft_si}).
This implies that the star formation rate 
in these regions is not only particularly low today, but has already been low in the past.
The lifetime of synchrotron-emitting electrons in a magnetic field strength of  $5$~$\mu$G (assuming equipartition between
relativistic particles and the magnetic field) observed at $4$~GHz 
is $\sim 45$~Myr. We suggest that ram pressure stripping as described in Vollmer (2003) and  Vollmer et al. (2008a) 
is responsible for the decreased star formation efficiencies in these regions. It is expected that this gas
is just leaving the disk. 
On the other hand, the regions of low star formation efficiency $SFE_{\rm tot}$ in NGC~4654 and NGC~4330 do not show detectable radio
continuum emission (Fig.~\ref{fig:sft_si}). This implies that star formation has always been small in these regions (NGC~4330) or that
the timescale at which star formation dropped significantly is longer than the characteristic time for the decline of synchrotron emission at $1.4$~GHz which 
is $\sim 75$~Myr (NGC~4654).
We note that in NGC~4501 and NGC~4654 the regions of decreased star formation efficiency rotate into the ram pressure wind.
The fifth galaxy with a low star formation efficiency with respect to the total gas in the extraplanar gas is NGC~4438 (Fig.~\ref{fig:n4438_sft_si}).
In this galaxy even the star formation efficiency with respect to the molecular gas is reduced by a factor of $\sim 3$ (see Sect.~\ref{sec:n4438})
with respect to the average molecular gas consumption time of this sample of $2.4 \pm 1.1$~Gyr (see Sect.~\ref{sec:results}).

On the other hand, the western gas arm of NGC~4569 (Fig.~\ref{fig:plot_n4569}), which consists of resettling gas (Vollmer et al. 2004a), 
has a star formation efficiency $SFE_{\rm tot}$ indistinguishable from that of the inner disk. Thus, the star formation efficiency in backfalling,
resettling gas is higher than that of actively stripped gas.

\subsection{The unusually low $SFE_{\rm mol}$ of NGC~4438 \label{sec:n4438}}

In galactic disks, the standard environment for star formation, it consistently appears that the star formation rate is proportional to the H$_2$ mass.
This consistency implies that the variation in the structure of the molecular ISM and in the stellar IMF is small. The low $SFE_{\rm mol}$ in the ram-pressure 
stripped extraplanar gas of NGC~4438 (Fig.~\ref{fig:rmol_graph}) suggests that the stripping mechanism has altered the state of the gas.  
Since the extraplanar molecular gas has been located within the optical radius before stripping (Vollmer et al. 2005b), its metallicity is expected
to be around the solar value. 
Thus we do not expect the $N({\rm H}_{2})/I_{\rm CO}$ ratio of this gas to be affected by an unusual metallicity.\footnote{However, given the peculiar 
environment and properties of the extraplanar gas, the $N({\rm H}_{2})/I_{\rm CO}$ ratio of this gas has to be regarded with caution.} 
It is interesting to note that in another non-disk environment, that of Tidal Dwarf 
Galaxies, $SFE_{\rm mol}$ is similar to that of spirals. The metallicity of Tidal Dwarf Galaxies is only slightly subsolar, such that a classical 
$N({\rm H}_{2})/I_{\rm CO}$ ratio probably applies (Braine et al. 2001).
In ellipticals, the situation appears similar albeit with considerably greater scatter (Lees et al. 1991).

What other cases of an unusually low $SFE_{\rm mol}$ are known?  Probably the best known case is that of the Taffy Galaxies (UGC 12914/5), an ISM-ISM collision discovered 
by Condon et al. (1993).  In the bridge region, there is roughly five times the CO emission of the entire Milky Way (Braine et al. 2003; Gao et al. 2003) but the 
star formation rate is only $\sim 1/10$th that of the Galaxy.  The other galaxy-scale ISM-ISM collision, found by Condon et al. (2002), the UGC 813/816 system, shows 
the same low $SFE_{\rm mol}$ (Braine et al. 2004).  It is also suspected that NGC~4438 had an ISM-ISM collision with M~86 (Kenney et al. 2008), and
possibly also NGC~4435 (Vollmer et al. 2005b).
The northern part of NGC~3227 in Arp 94 also shows abundant H$_2$ with little or no star formation (Lisenfeld et al. 2008).  
Some parts of Stephan's Quintet may also fall into this category.

What are the common features of these systems?
(i) The CO lines are broad, broader than for disk star forming regions. (ii) Strong radio continuum emission is present (see Condon et al. 1993, 2002 for the Taffy 
and UGC 813/6 systems, Kotanyi et al. 1983, Vollmer et al. 2009 for NGC 4438, and Mundell et al. 1995 for Arp 94).  (iii) Except for Arp 94 where the situation is unclear, 
the other systems involve gas-gas collisions and thus shocks.  (iv) Except for Arp 94 where the situation is unclear, in each case the region with a low 
$SFE_{\rm mol}$ was created only a few 10~Myr ago.

\subsection{The influence of ram pressure on the ISM and star formation}

We found that for all but one galaxy the cluster environment does not affect the star formation efficiency with respect to the molecular gas.
Gas truncation is not associated with major changes in the total gas surface density distribution of the inner disk of Virgo spiral galaxies.
In three galaxies (NGC~4430, NGC~4501, and NGC~4522) the molecular fraction is increased by a factor of $1.5$ to $2$ on the windward side 
of the galactic disk. The star formation efficiency with respect to the total gas is increased in the asymmetric regions
of high molecular fractions in NGC~4330 and NGC~4522, but not in NGC~4501.
A significant increase of the star formation efficiency with respect to the molecular gas content on the windward side of ram 
pressure-stripped galaxies is not observed. 
The ram-pressure stripped extraplanar gas of 3 highly inclined spiral galaxies (NGC~4330, NGC~4438, and NGC~4522) shows a depressed star formation 
efficiency with respect to the total gas, and one of them (NGC~4438) shows a depressed rate even with respect to the molecular gas. 
Ram pressure stripping thus generally decreases the star formation rate.

How does ram pressure physically affect the ISM and star formation? On the one hand, ram pressure compresses the gas leading
to higher volume densities. Higher gas densities lead to higher molecular fractions and star formation rates.
On the other hand, transfer of energy between the intracluster medium and the ISM is also expected.
Before and during momentum transfer by ram pressure, shocks are driven into the ISM which increase its thermal
and turbulent pressure. The shock-heated molecular gas can be detected in the pure rotational H$_{2}$ transitions S(0) to S(7) 
(Sivanandam et al. 2010, Wong et al. 2012b in prep.).
In the absence of the confining gravitational potential of the galactic disk, and as it experiences a diverging gas flow in the earlier stages of ram
pressure, the stripped gas thus disperses, its density decreases, star formation drops significantly, and the star formation efficiency is low. On the 
other hand, the star formation efficiency is high in backfalling gas that re-settles in the galactic disk (NGC~4569).  Gas stripped from the disk which 
never achieves escape velocity and is therefore not dispersed in early stages of stripping may resettle back into the disk. Dynamical simulations indicate 
that the overdense gas arms which can form in stripped gas (Vollmer et al. 2008b) resist dispersal better than lower density gas, and are more likely 
to be part of the backfalling gas. Such backfalling extraplanar arms, like the one in NGC~4569, are denser than the average extraplanar gas in early 
stage stripping, plus may experiencing a converging rather than diverging flow. These factors may account for the higher star formation efficiency in 
the extraplanar gas of NGC~4569. For the moment we know very little about the ratio between momentum and energy transfer during a ram pressure
stripping event.

\section{Conclusions \label{sec:conclusions}}

We have compared the 2-dimensional spatial distributions of gas and star formation efficiency in 12 Virgo spirals using 
VIVA VLA H{\sc i} (Chung et al. 2009), Nobeyama 45m CO(1-0) (Kuno et al. 2007), IRAM 30m CO(2-1) (Vollmer et al. 2005b, Vollmer et al. 2008b,
Vollmer et al. 2011), GALEX UV (Gil de Paz et al. 2007), and SPITSOV Spitzer infrared (Wong et al. 2012a, in prep.) data.
The star formation rates were derived from the UV and total infrared (TIR) emission based on the
8, 24, 70, and 160~$\mu$m data. To gain spatial 
resolution we made the approximation $I_{160\mu{\rm m}} \sim I_{70\mu{\rm m}}$ for most of the galaxies. This TIR emission
was compared to the approximation only based on the $24$~$\mu$m data used in the literature.
As expected, the latter recipe underestimates the TIR in the outer parts of the disk where the
stellar radiation field is weak and  dust temperatures are low. In the galaxy centers of high
gas density, the description based on $24$~$\mu$m data overestimates the TIR emission up to a factor of $2$-$3$.

Our sample of Virgo spiral galaxies contains 5 galaxies which are or were affected by ram pressure stripping,
one galaxy which underwent a tidal interaction, and two galaxies which had a tidal interaction and are
now undergoing ram pressure stripping. 
Six galaxies show strongly truncated gas disks inside the optical radius. Two galaxies have anemic starforming disks,
a type of galaxy which is also found outside clusters.
The gas distributions and star formation efficiencies were compared to individual galaxies of the THINGS (Walter et al. 2008) and
HERACLES (Leroy et al. 2009) samples of nearby spiral galaxies.

Based on our analysis we conclude that
\begin{enumerate}
\item
Inside the truncation radius, the inner disk molecular and atomic gas distributions of truncated Virgo spirals are not significantly different 
from those of nearby field galaxies.
Gas truncation is not associated with major changes in the gas surface density of the inner disk of Virgo spiral galaxies.
\item
In three galaxies we detect a $1.5$--$2$ times enhanced molecular fraction at the windward sides of the galactic disks with respect to the leeward sides.
This interpretation of a high molecular fraction due to ram pressure compression is hampered by
the extraplanar gas located at the leeward side of the galactic disk in one edge-on galaxy.
\item
Except for NGC~4438, the cluster environment does not affect the star formation efficiency with respect to the molecular gas
confirming the results of Fumagalli et al. (2008).
As in nearby field spiral galaxies, the star formation rate is approximately proportional to the molecular gas
surface density (Bigiel et al. 2008, Leroy et al. 2008, Bigiel et al. 2011).
\item
The star formation efficiency with respect to the molecular gas of the extraplanar region in NGC~4438 is 
reduced by a factor of $\sim 3$ compared to a molecular gas consumption rate of $2$~Gyr. Displaced molecular gas can
thus yield a significantly reduced star formation efficiency $SFE_{\rm mol}$.
\item
We observe a possible moderate increase of the molecular fraction on the windward side of the ram pressure-stripped galaxies NGC~4330, NGC~4522,
and NGC~4501. The first two galaxies show an increased star formation efficiency with respect to the total gas in these regions. 
These effects are not seen in NGC~4569 and NGC~4654, which are also affected by ram pressure.
\item
All Virgo galaxies have counterparts for inner disk H{\sc i} and CO distributions
in the sample of nearby field galaxies. We propose that high star formation efficiencies with respect to the total gas at low total gas surface densities
can be explained by a higher stellar surface density. In these galaxies the star formation efficiency with 
respect to the molecular gas is normal.
\item
The ram-pressure stripped extraplanar gas of the edge-on spiral galaxies NGC~4330, NGC~4438, and NGC~4522 shows a depressed star formation efficiency 
with respect to the total gas surface density. The interpretation for NGC~4330 and NGC~4522 is that stripped gas loses the gravitational confinement 
of the galactic disk, plus the gas flow is diverging, thus the gas density decreases and star formation drops.
\item
In addition to NGC~4330 and NGC~4522, we found two such regions of low star formation efficiency in the more face-on galaxies NGC~4501 
and NGC~4654 which both are undergoing ram pressure stripping. All regions with a low star formation efficiency also show
low radio continuum emission or an unusually steep spectral index between $1.4$ and $4.8$~GHz.
\item
On the other hand, the stripped extraplanar gas of NGC~4569 has a normal star formation efficiency with respect to the total gas. 
It differs from the galaxies in the earlier stages of stripping, perhaps because its extraplanar gas flow is converging as the gas resettles back into the disk.
\end{enumerate}

We do not find evidence of a strongly increased star formation efficiency due to the action of ram pressure.
The most significant effect of ram pressure stripping on the star formation of the galaxies in our sample is a strongly 
decreased star formation efficiency with respect to the total gas surface density in the stripped gas.
This is important since it implies that even though some stars form in ram pressure stripped gas, the likely fate for the vast majority 
of the stripped gas is to be heated and join the general intracluster medium.

\begin{acknowledgements}
This research has made use of the GOLDMine Database.
This work has also been supported by the National Research Foundation of Korea grant 2011-8-0993 and 2011-8-1679, and Yonsei research grant 2011-1-0096.
\end{acknowledgements}

\appendix

\section{CO moment 0 maps}

\begin{figure*}
\centering
  \includegraphics[width=14cm]{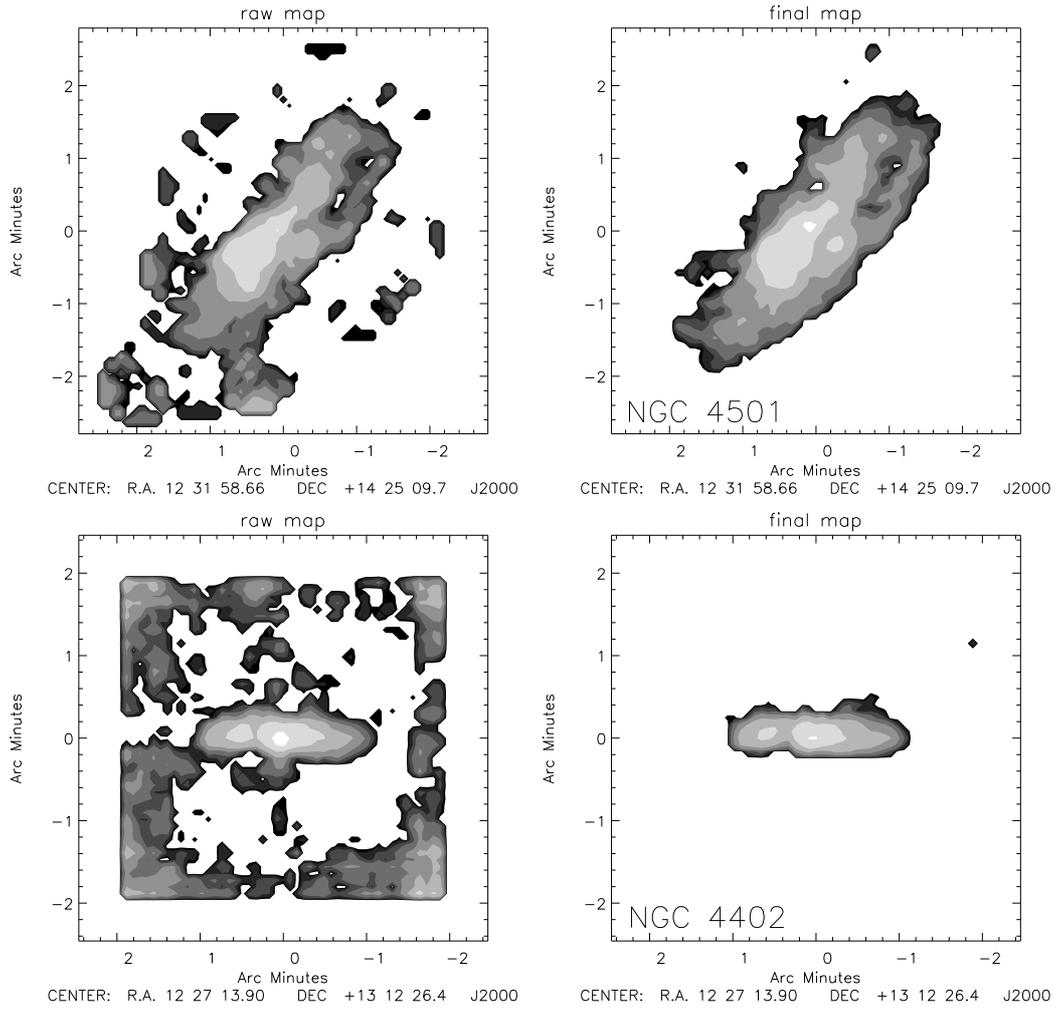}
  \caption{CO moment 0 maps of NGC~4501 and NGC~4402. Left panels: resulting from clipping the CO
    data cube at a constant level of $75$~mK. Right panels: after the procedure described in Sect.~\ref{sec:gasdens}
    involving the H{\sc i} data cubes.
  \label{fig:compareCO}}%
\end{figure*}

\section{The star formation efficiency of the sample galaxies}

\begin{figure*}
  \centering
  \includegraphics[width=14cm]{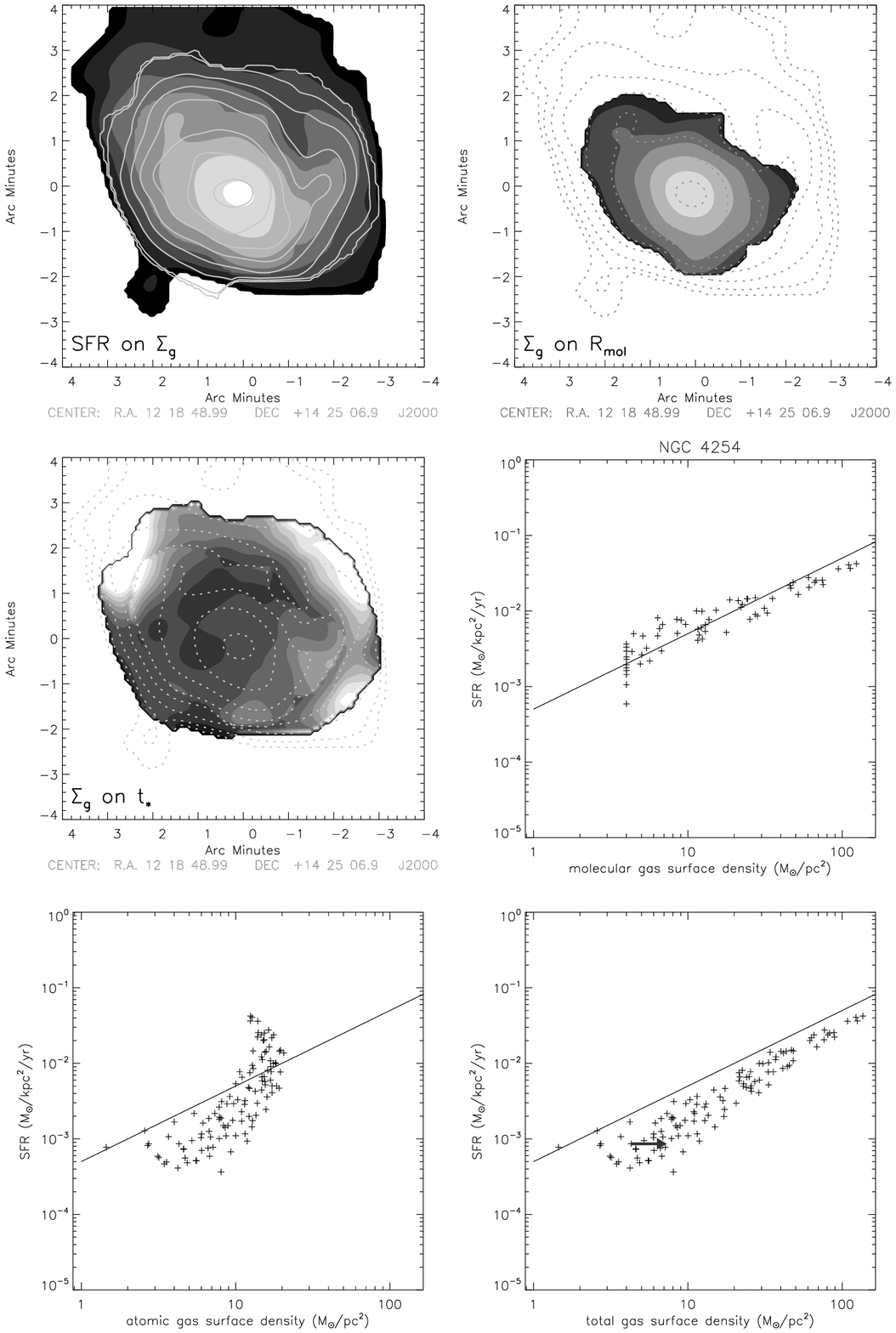}
  \caption{NGC~4254: top left: star formation rate (from UV+TIR) in contours on a greyscale map of the total gas surface density.
    Contour levels are $(1,2,4,8,16,32,64,128)\ 3 \times 10^{-4}$~M$_{\odot}$kpc$^{-2}$pc$^{-1}$.
    Greyscale levels are $(1,2,4,8,16,32,64,128)$~M$_{\odot}$pc$^{-2}$. Top right: 
    Total gas surface density contours on the molecular gas fraction $\Sigma_{\rm H_{2}}/\Sigma_{\rm HI}$.
    Greyscale levels are $(1,2,4,8,16,32,64) \times 0.1$.
    Middle left: total gas surface density contours on the star formation timescale $t_{*}=\Sigma_{\rm g}/\dot{\Sigma}_{*}$.
    Greyscale levels are $(1,10,20,30,40,50,60,70,80,90,100)$~10~Myr.
    Middle right: star formation rate $\dot{\Sigma}_{*}$ as a function of the molecular gas surface density
    $\Sigma_{\rm H_{2}}$. The vertical set of points at the lowest $\Sigma_{\rm mol}$ are upper limits for $\Sigma_{\rm mol}$. 
    Lower left: star formation rate $\dot{\Sigma}_{*}$ as a function of the atomic gas surface density
    $\Sigma_{\rm HI}$. Lower right: star formation rate $\dot{\Sigma}_{*}$ as a function of the total gas surface density
    $\Sigma_{\rm g}=\Sigma_{\rm H_{2}}+\Sigma_{\rm HI}$. The solid line corresponds to a star formation timescale of
    $t_{*}=2$~Gyr. The arrow takes into account the CO detection limit.
  \label{fig:plot_n4254}}%
\end{figure*}

\begin{figure*}
  \centering
  \includegraphics[width=14cm]{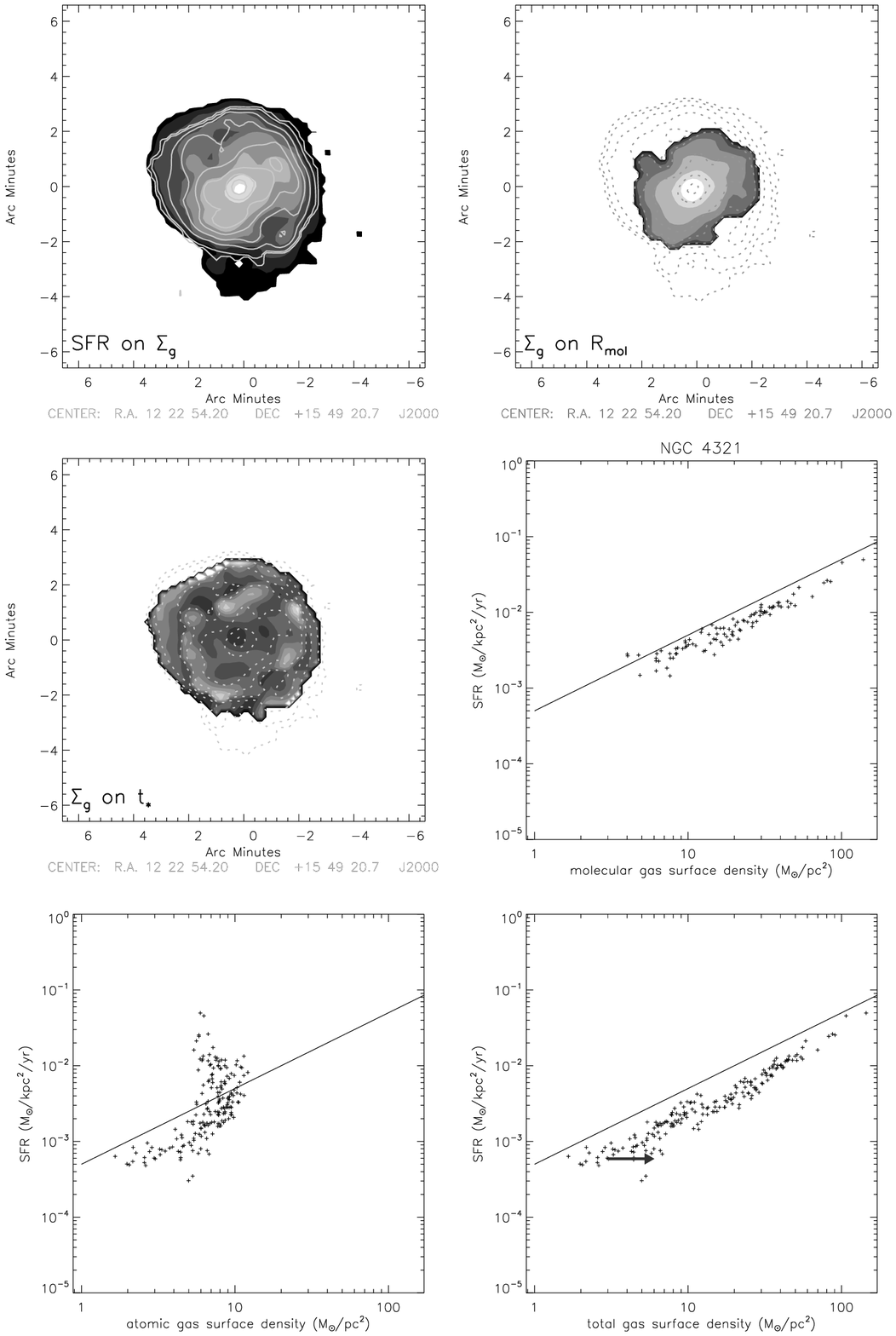}
  \caption{NGC~4321: top left: star formation rate (from UV+TIR) in contours on a greyscale map of the total gas surface density.
    Contour levels are $(1,2,4,8,16,32,64,128)\ 3 \times 10^{-4}$~M$_{\odot}$kpc$^{-2}$pc$^{-1}$.
    Greyscale levels are $(1,2,4,8,16,32,64,128)$~M$_{\odot}$pc$^{-2}$. Top right: 
    Total gas surface density contours on the molecular gas fraction $\Sigma_{\rm H_{2}}/\Sigma_{\rm HI}$.
    Greyscale levels are $(1,2,4,8,16,32,64,128) \times 0.1$.
    Middle left: total gas surface density contours on the star formation timescale $t_{*}=\Sigma_{\rm g}/\dot{\Sigma}_{*}$.
    Greyscale levels are $(1,10,20,30,40,50,60,70,80,90,100) \times 10$~Myr.
    Middle right: star formation rate $\dot{\Sigma}_{*}$ as a function of the molecular gas surface density
    $\Sigma_{\rm H_{2}}$. Lower left: star formation rate $\dot{\Sigma}_{*}$ as a function of the atomic gas surface density
    $\Sigma_{\rm HI}$. Lower right: star formation rate $\dot{\Sigma}_{*}$ as a function of the total gas surface density
    $\Sigma_{\rm g}=\Sigma_{\rm H_{2}}+\Sigma_{\rm HI}$. The solid line corresponds to a star formation timescale of
    $t_{*}=2$~Gyr. The arrow takes into account the CO detection limit.
  \label{fig:plot_n4321}}
\end{figure*}

\begin{figure*}
  \centering
  \includegraphics[width=14cm]{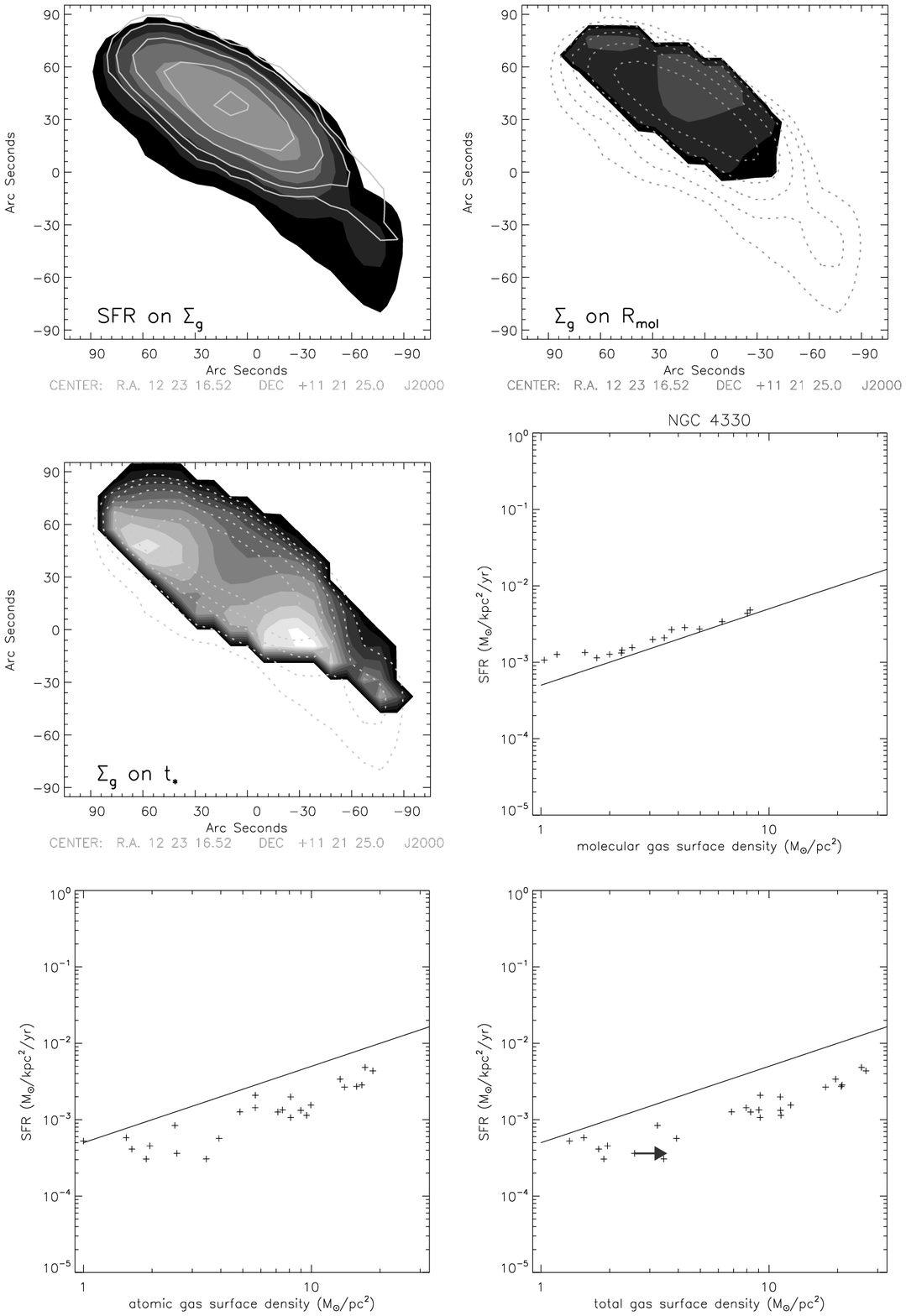}
  \caption{NGC~4330: top left: star formation rate (from UV+TIR) in contours on a greyscale map of the total gas surface density.
    Contour levels are $(1,2,4,8,16)\ 3 \times 10^{-4}$~M$_{\odot}$kpc$^{-2}$pc$^{-1}$.
    Greyscale levels are $(1,2,4,8,16)$~M$_{\odot}$pc$^{-2}$. Top right: 
    Total gas surface density contours on the molecular gas fraction $\Sigma_{\rm H_{2}}/\Sigma_{\rm HI}$.
    Greyscale levels are $(1,2,4) \times 0.1$.
    Middle left: total gas surface density contours on the star formation timescale $t_{*}=\Sigma_{\rm g}/\dot{\Sigma}_{*}$.
    Greyscale levels are $(1,10,20,30,40,50,60,70,80,90,100) \times 10$~Myr.
    Middle right: star formation rate $\dot{\Sigma}_{*}$ as a function of the molecular gas surface density
    $\Sigma_{\rm H_{2}}$. Lower left: star formation rate $\dot{\Sigma}_{*}$ as a function of the atomic gas surface density
    $\Sigma_{\rm HI}$. Lower right: star formation rate $\dot{\Sigma}_{*}$ as a function of the total gas surface density
    $\Sigma_{\rm g}=\Sigma_{\rm H_{2}}+\Sigma_{\rm HI}$. The solid line corresponds to a star formation timescale of
    $t_{*}=2$~Gyr. The arrow takes into account the CO detection limit.
  \label{fig:plot_n4330}}%
\end{figure*}

\begin{figure*}
  \centering
  \includegraphics[width=14cm]{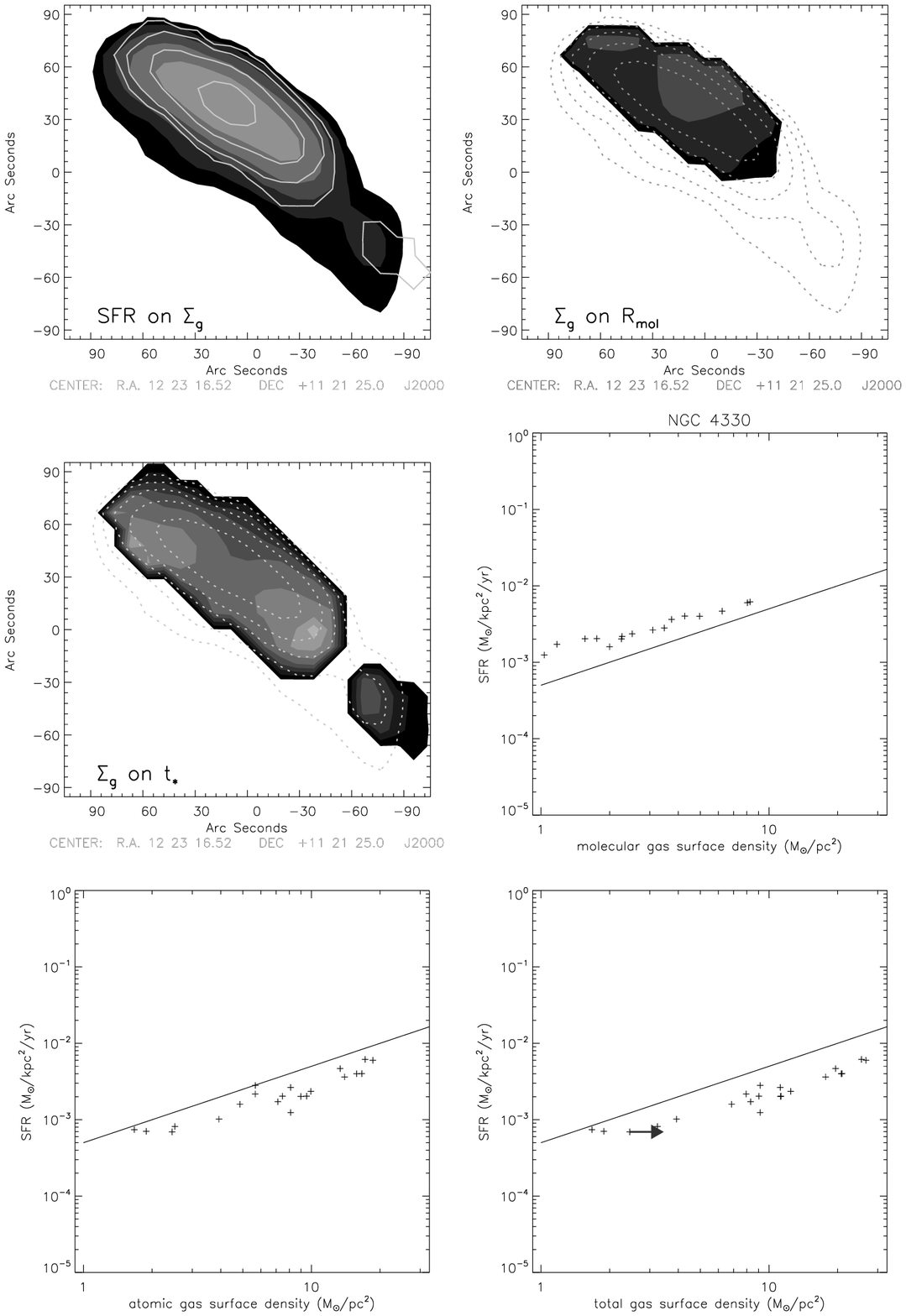}
  \caption{NGC~4330: top left: star formation rate based on H$\alpha$ and TIR emission in contours on a greyscale map of 
    the total gas surface density.
    Contour levels are $(1,2,4,8,16)\ 5 \times 10^{-4}$~M$_{\odot}$kpc$^{-2}$pc$^{-1}$.
    Greyscale levels are $(1,2,4,8,16)$~M$_{\odot}$pc$^{-2}$. Top right: 
    Total gas surface density contours on the molecular gas fraction $\Sigma_{\rm H_{2}}/\Sigma_{\rm HI}$.
    Greyscale levels are $(1,2,4) \times 0.1$.
    Middle left: total gas surface density contours on the star formation timescale $t_{*}=\Sigma_{\rm g}/\dot{\Sigma}_{*}$.
    Greyscale levels are $(1,10,20,30,40,50,60,70,80,90,100) \times 10$~Myr.
    Middle right: star formation rate $\dot{\Sigma}_{*}$ as a function of the molecular gas surface density
    $\Sigma_{\rm H_{2}}$. Lower left: star formation rate $\dot{\Sigma}_{*}$ as a function of the atomic gas surface density
    $\Sigma_{\rm HI}$. Lower right: star formation rate $\dot{\Sigma}_{*}$ as a function of the total gas surface density
    $\Sigma_{\rm g}=\Sigma_{\rm H_{2}}+\Sigma_{\rm HI}$. The solid line corresponds to a star formation timescale of
    $t_{*}=2$~Gyr. The arrow takes into account the CO detection limit.
  \label{fig:plot_n4330_ha}}%
\end{figure*}

\begin{figure*}
  \centering
  \includegraphics[width=14cm]{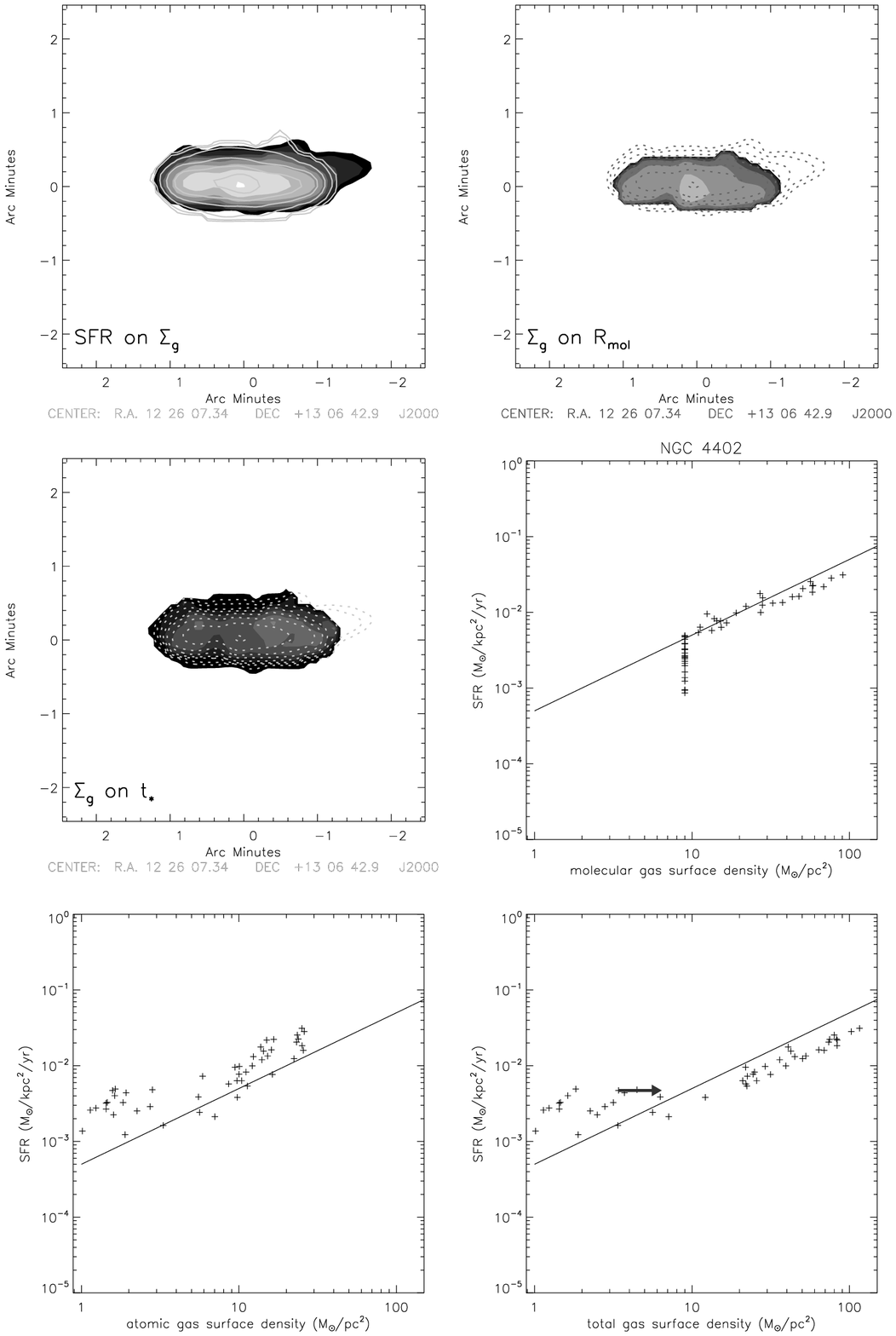}
  \caption{NGC~4402: top left: star formation rate (from UV+TIR) in contours on a greyscale map of the total gas surface density.
    Contour levels are $(1,2,4,8,16,32)\ 8 \times 10^{-4}$~M$_{\odot}$kpc$^{-2}$pc$^{-1}$.
    Greyscale levels are $(1,2,4,8,16,32,64,128)$~M$_{\odot}$pc$^{-2}$. Top right: 
    Total gas surface density contours on the molecular gas fraction $\Sigma_{\rm H_{2}}/\Sigma_{\rm HI}$.
    Greyscale levels are $(1,2,4,8,16,32) \times 0.1$.
    Middle left: total gas surface density contours on the star formation timescale $t_{*}=\Sigma_{\rm g}/\dot{\Sigma}_{*}$.
    Greyscale levels are $(1,10,20,30,40,50) \times 10$~Myr.
    Middle right: star formation rate $\dot{\Sigma}_{*}$ as a function of the molecular gas surface density
    $\Sigma_{\rm H_{2}}$. The vertical set of points at the lowest $\Sigma_{\rm mol}$ are upper limits for $\Sigma_{\rm mol}$.
    Lower left: star formation rate $\dot{\Sigma}_{*}$ as a function of the atomic gas surface density
    $\Sigma_{\rm HI}$. Lower right: star formation rate $\dot{\Sigma}_{*}$ as a function of the total gas surface density
    $\Sigma_{\rm g}=\Sigma_{\rm H_{2}}+\Sigma_{\rm HI}$. The solid line corresponds to a star formation timescale of
    $t_{*}=2$~Gyr. The arrow takes into account the CO detection limit.
  \label{fig:plot_n4402}}%
\end{figure*}

\begin{figure*}
  \centering
  \includegraphics[width=14cm]{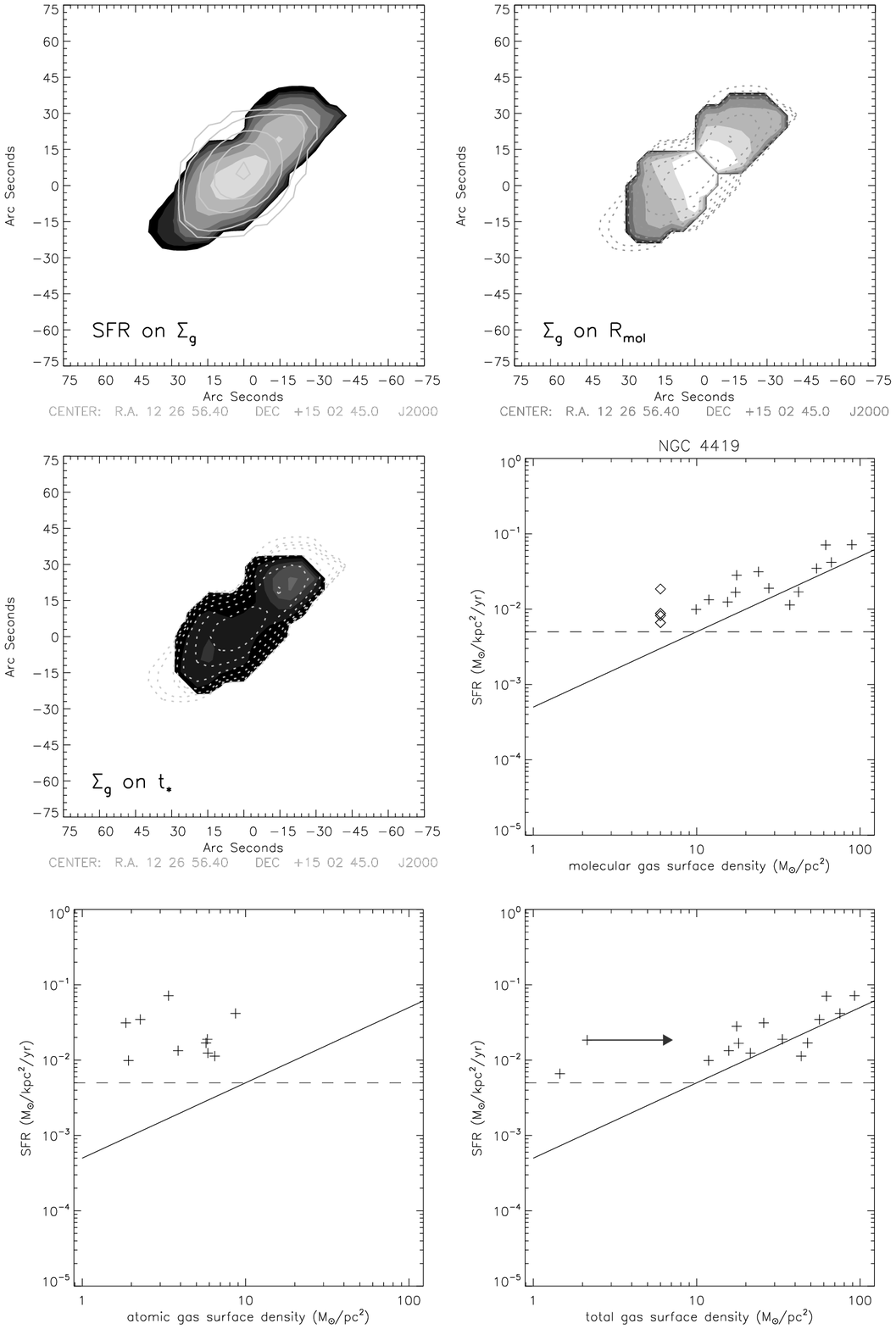}
  \caption{NGC~4419: top left: star formation rate (from UV+TIR) in contours on a greyscale map of the total gas surface density.
    Contour levels are $(1,2,4,8,16,32)\ 5 \times 10^{-3}$~M$_{\odot}$kpc$^{-2}$pc$^{-1}$.
    Greyscale levels are $(1,2,4,8,16,32,64)$~M$_{\odot}$pc$^{-2}$. Top right: 
    Total gas surface density contours on the molecular gas fraction $\Sigma_{\rm H_{2}}/\Sigma_{\rm HI}$.
    Greyscale levels are $(1,2,4,8,16,32,64,128) \times 0.1$.
    Middle left: total gas surface density contours on the star formation timescale $t_{*}=\Sigma_{\rm g}/\dot{\Sigma}_{*}$.
    Greyscale levels are $(1,10,20,30,40) \times 10$~Myr.
    Middle right: star formation rate $\dot{\Sigma}_{*}$ as a function of the molecular gas surface density
    $\Sigma_{\rm H_{2}}$. Diamonds are upper limits for $\Sigma_{\rm mol}$.
    Lower left: star formation rate $\dot{\Sigma}_{*}$ as a function of the atomic gas surface density
    $\Sigma_{\rm HI}$. Lower right: star formation rate $\dot{\Sigma}_{*}$ as a function of the total gas surface density
    $\Sigma_{\rm g}=\Sigma_{\rm H_{2}}+\Sigma_{\rm HI}$. The solid line corresponds to a star formation timescale of
    $t_{*}=2$~Gyr. The dashed line indicates the level at which we had to clip
    the star formation rate due to PSF effects of the bright nucleus. The arrow takes into account the CO detection limit.
  \label{fig:plot_n4419}}%
\end{figure*}

\begin{figure*}
  \centering
  \includegraphics[width=14cm]{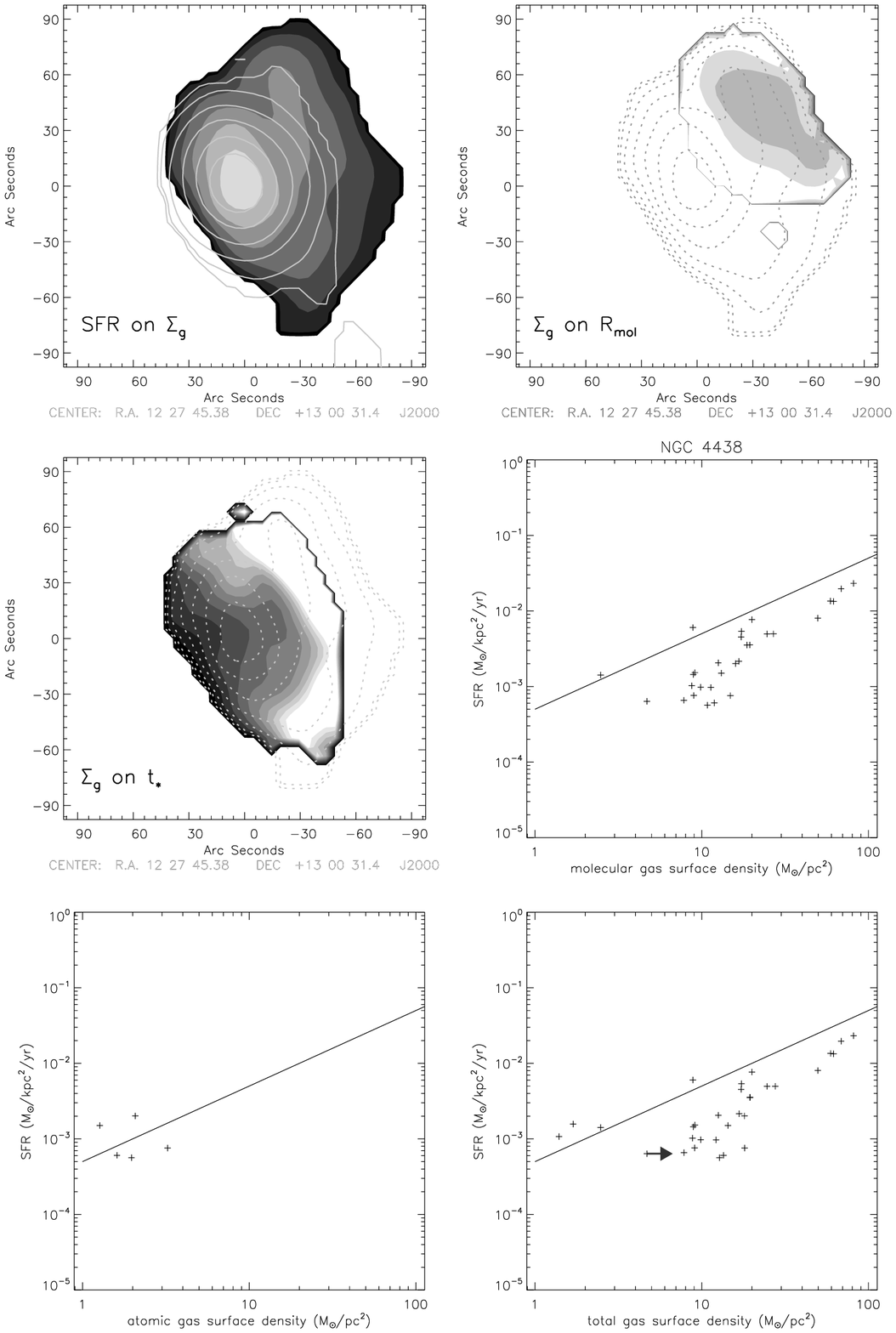}
  \caption{NGC~4438: top left: star formation rate (from UV+TIR) in contours on a greyscale map of the total gas surface density.
    Contour levels are $(1,2,4,8,16,32)\ 4.5 \times 10^{-4}$~M$_{\odot}$kpc$^{-2}$pc$^{-1}$.
    Greyscale levels are $(1,2,4,8,16,32,64)$~M$_{\odot}$pc$^{-2}$. Top right: 
    Total gas surface density contours on the molecular gas fraction $\Sigma_{\rm H_{2}}/\Sigma_{\rm HI}$.
    Greyscale levels are $(1,2,4,8,16,32,64,128) \times 0.1$.
    Middle left: total gas surface density contours on the star formation timescale $t_{*}=\Sigma_{\rm g}/\dot{\Sigma}_{*}$.
    Greyscale levels are $(1,10,20,30,40) \times 10$~Myr.
    Middle right: star formation rate $\dot{\Sigma}_{*}$ as a function of the molecular gas surface density
    $\Sigma_{\rm H_{2}}$. Lower left: star formation rate $\dot{\Sigma}_{*}$ as a function of the atomic gas surface density
    $\Sigma_{\rm HI}$. Lower right: star formation rate $\dot{\Sigma}_{*}$ as a function of the total gas surface density
    $\Sigma_{\rm g}=\Sigma_{\rm H_{2}}+\Sigma_{\rm HI}$. The solid line corresponds to a star formation timescale of
    $t_{*}=2$~Gyr. The dashed line indicates the level at which we had to clip
    the star formation rate due to PSF effects of the bright nucleus. The arrow takes into account the CO detection limit.
  \label{fig:plot_n4438}}%
\end{figure*}

\begin{figure*}
  \centering
  \includegraphics[width=14cm]{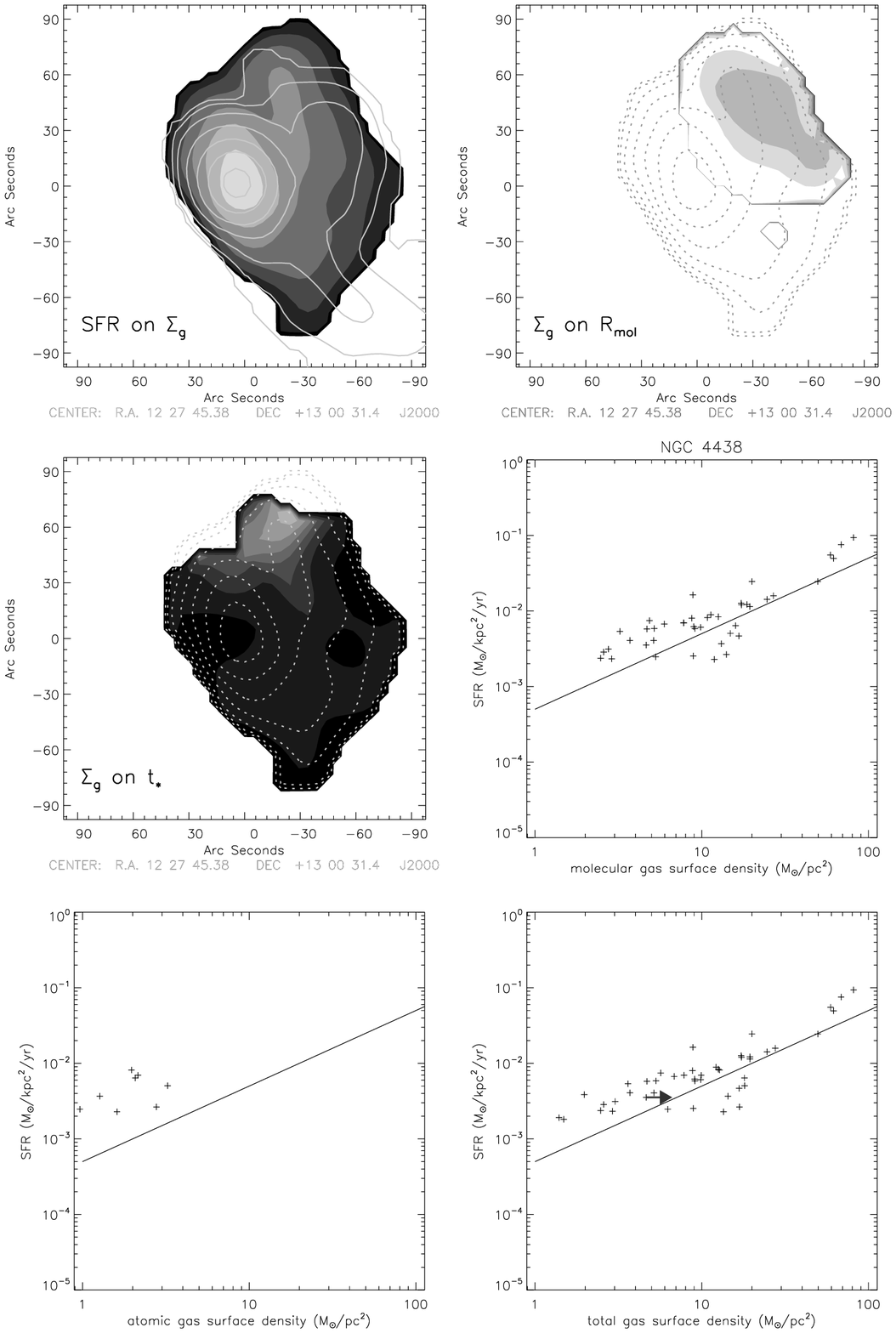}
  \caption{NGC~4438: top left: star formation rate based on H$\alpha$ and TIR emission in contours on a greyscale map of 
    the total gas surface density.
    Contour levels are $(1,2,4,8,16,32)\ 15 \times 10^{-4}$~M$_{\odot}$kpc$^{-2}$pc$^{-1}$.
    Greyscale levels are $(1,2,4,8,16,32)$~M$_{\odot}$pc$^{-2}$. Top right: 
    Total gas surface density contours on the molecular gas fraction $\Sigma_{\rm H_{2}}/\Sigma_{\rm HI}$.
    Greyscale levels are $(1,2,4,8) \times 0.1$.
    Middle left: total gas surface density contours on the star formation timescale $t_{*}=\Sigma_{\rm g}/\dot{\Sigma}_{*}$.
    Greyscale levels are $(1,10,20,30,40,50,60,70,80,90,100) \times 10$~Myr.
    Middle right: star formation rate $\dot{\Sigma}_{*}$ as a function of the molecular gas surface density
    $\Sigma_{\rm H_{2}}$. Lower left: star formation rate $\dot{\Sigma}_{*}$ as a function of the atomic gas surface density
    $\Sigma_{\rm HI}$. Lower right: star formation rate $\dot{\Sigma}_{*}$ as a function of the total gas surface density
    $\Sigma_{\rm g}=\Sigma_{\rm H_{2}}+\Sigma_{\rm HI}$. The solid line corresponds to a star formation timescale of
    $t_{*}=2$~Gyr. The arrow takes into account the CO detection limit.
  \label{fig:plot_n4438_ha}}%
\end{figure*}

\begin{figure*}
  \centering
  \includegraphics[width=14cm]{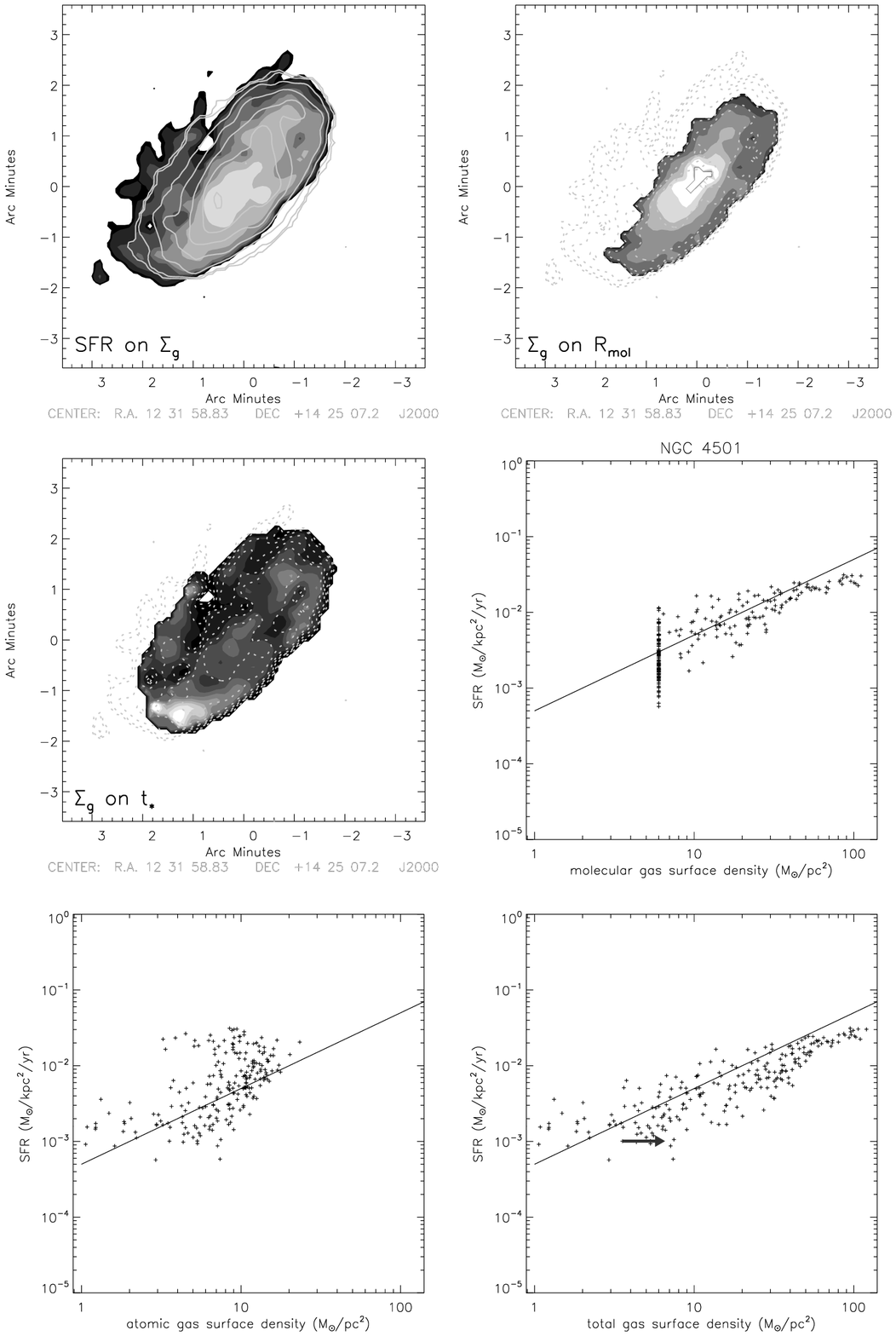}
  \caption{NGC~4501: top left: star formation rate (from UV+TIR) in contours on a greyscale map of the total gas surface density.
    Contour levels are $(1,2,4,8,16,32,64)\ 5 \times 10^{-4}$~M$_{\odot}$kpc$^{-2}$pc$^{-1}$.
    Greyscale levels are $(1,2,4,8,16,32,64)$~M$_{\odot}$pc$^{-2}$. Top right: 
    Total gas surface density contours on the molecular gas fraction $\Sigma_{\rm H_{2}}/\Sigma_{\rm HI}$.
    Greyscale levels are $(1,2,4,8,16,32,64,128) \times 0.1$.
    Middle left: total gas surface density contours on the star formation timescale $t_{*}=\Sigma_{\rm g}/\dot{\Sigma}_{*}$.
    Greyscale levels are $(1,10,20,30,40,50,60,70,80,90,100) \times 10$~Myr.
    Middle right: star formation rate $\dot{\Sigma}_{*}$ as a function of the molecular gas surface density
    $\Sigma_{\rm H_{2}}$. The vertical set of points at the lowest $\Sigma_{\rm mol}$ are upper limits for $\Sigma_{\rm mol}$.
    Lower left: star formation rate $\dot{\Sigma}_{*}$ as a function of the atomic gas surface density
    $\Sigma_{\rm HI}$. Lower right: star formation rate $\dot{\Sigma}_{*}$ as a function of the total gas surface density
    $\Sigma_{\rm g}=\Sigma_{\rm H_{2}}+\Sigma_{\rm HI}$. The solid line corresponds to a star formation timescale of
    $t_{*}=2$~Gyr. The arrow takes into account the CO detection limit.
  \label{fig:plot_n4501}}%
\end{figure*}

\begin{figure*}
  \centering
  \includegraphics[width=14cm]{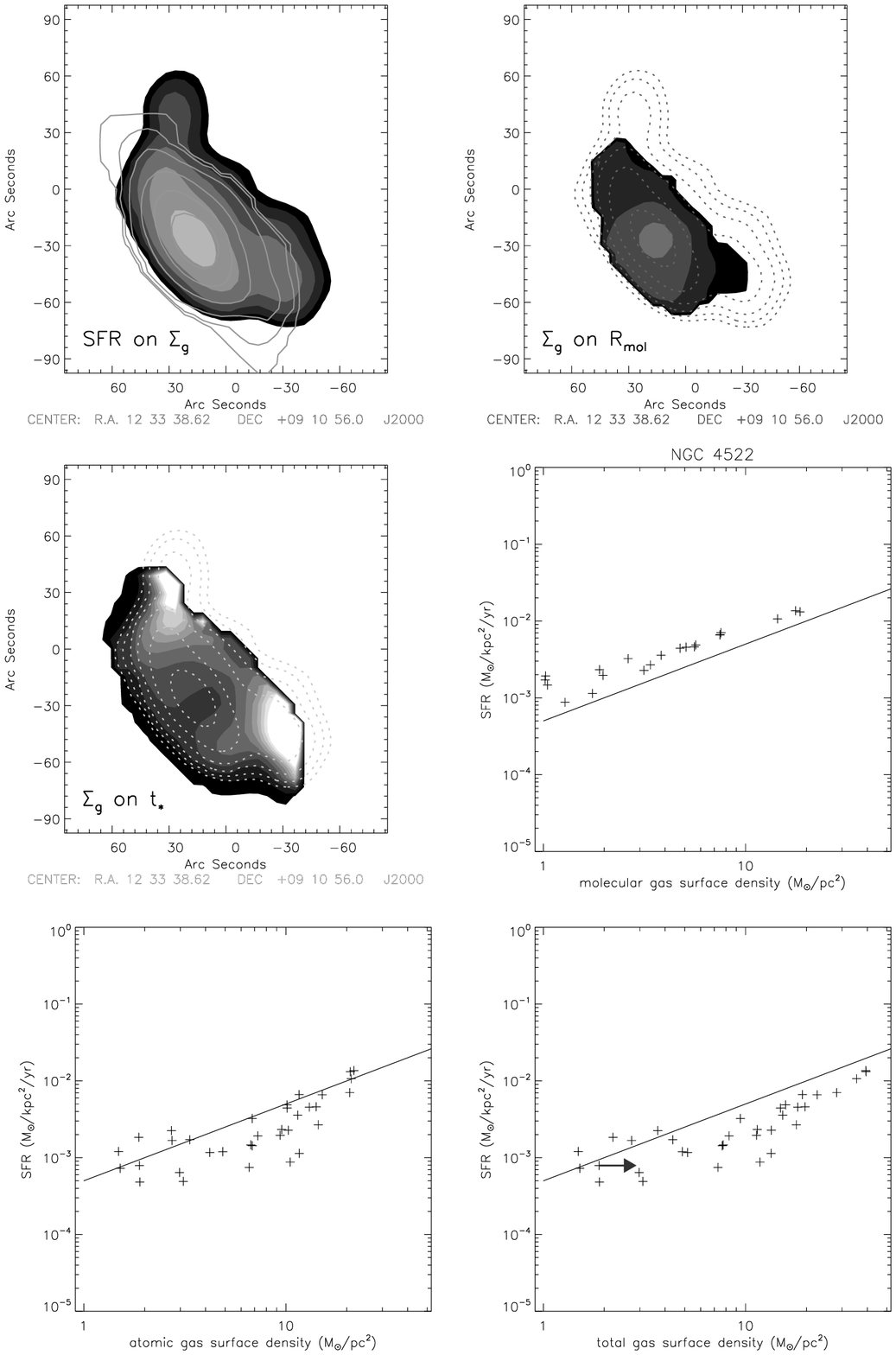}
  \caption{NGC~4522: top left: star formation rate (from UV+TIR) in contours on a greyscale map of the total gas surface density.
    Contour levels are $(1,2,4,8,16,32)\ 3 \times 10^{-4}$~M$_{\odot}$kpc$^{-2}$pc$^{-1}$.
    Greyscale levels are $(1,2,4,8,16,32)$~M$_{\odot}$pc$^{-2}$. Top right: 
    Total gas surface density contours on the molecular gas fraction $\Sigma_{\rm H_{2}}/\Sigma_{\rm HI}$.
    Greyscale levels are $(1,2,4,8) \times 0.1$.
    Middle left: total gas surface density contours on the star formation timescale $t_{*}=\Sigma_{\rm g}/\dot{\Sigma}_{*}$.
    Greyscale levels are $(1,10,20,30,40,50,60,70,80,90,100) \times 10$~Myr.
    Middle right: star formation rate $\dot{\Sigma}_{*}$ as a function of the molecular gas surface density
    $\Sigma_{\rm H_{2}}$. Lower left: star formation rate $\dot{\Sigma}_{*}$ as a function of the atomic gas surface density
    $\Sigma_{\rm HI}$. Lower right: star formation rate $\dot{\Sigma}_{*}$ as a function of the total gas surface density
    $\Sigma_{\rm g}=\Sigma_{\rm H_{2}}+\Sigma_{\rm HI}$. The solid line corresponds to a star formation timescale of
    $t_{*}=2$~Gyr. The arrow takes into account the CO detection limit.
  \label{fig:plot_n4522}}%
\end{figure*}
 
\begin{figure*}
  \centering
  \includegraphics[width=14cm]{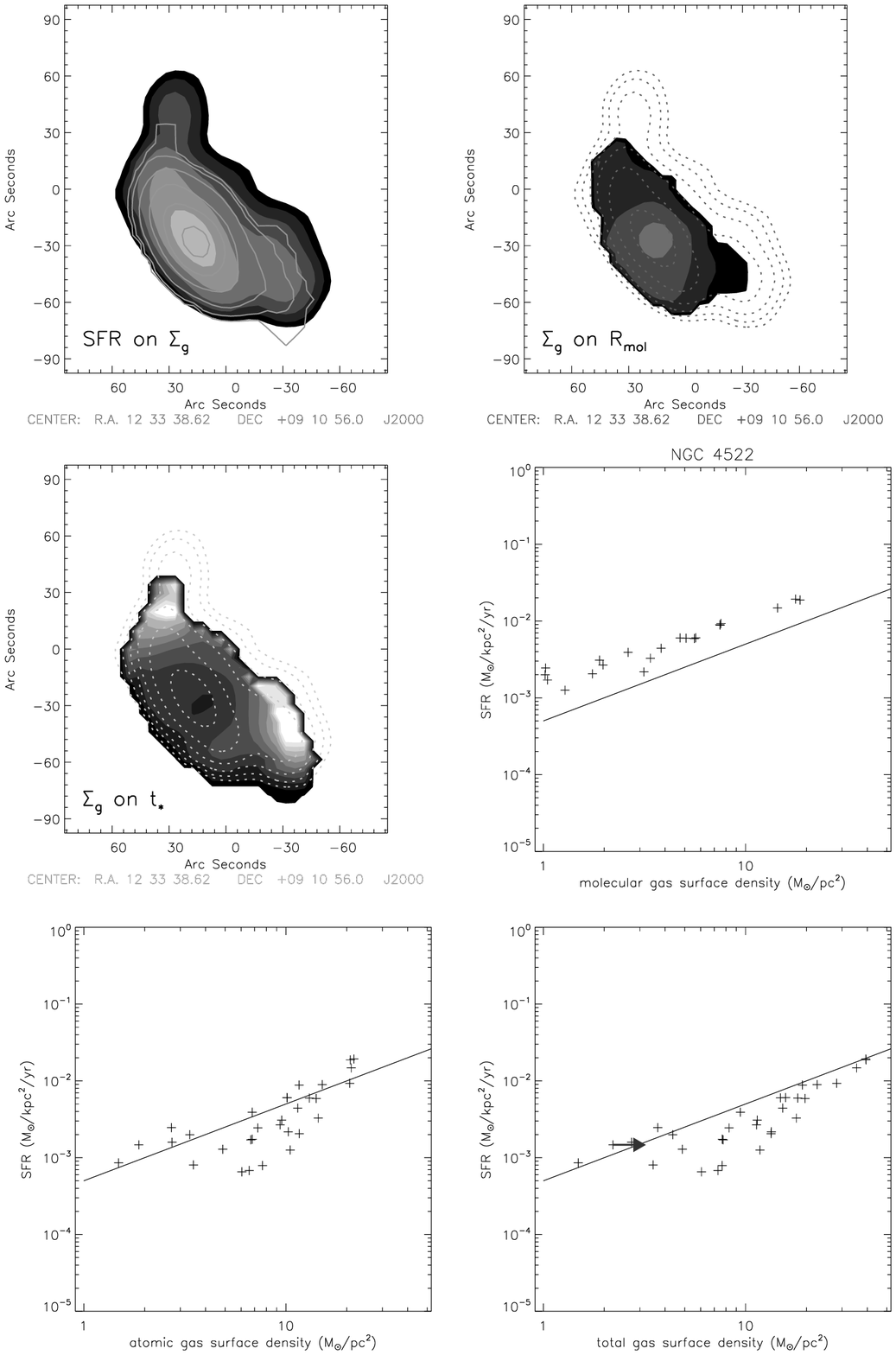}
  \caption{NGC~4522: top left: star formation rate based on H$\alpha$ and TIR emission in contours on a greyscale map of 
    the total gas surface density.
    Contour levels are $(1,2,4,8,16,32)\ 6 \times 10^{-4}$~M$_{\odot}$kpc$^{-2}$pc$^{-1}$.
    Greyscale levels are $(1,2,4,8,16,32)$~M$_{\odot}$pc$^{-2}$. Top right: 
    Total gas surface density contours on the molecular gas fraction $\Sigma_{\rm H_{2}}/\Sigma_{\rm HI}$.
    Greyscale levels are $(1,2,4,8) \times 0.1$.
    Middle left: total gas surface density contours on the star formation timescale $t_{*}=\Sigma_{\rm g}/\dot{\Sigma}_{*}$.
    Greyscale levels are $(1,10,20,30,40,50,60,70,80,90,100) \times 10$~Myr.
    Middle right: star formation rate $\dot{\Sigma}_{*}$ as a function of the molecular gas surface density
    $\Sigma_{\rm H_{2}}$. Lower left: star formation rate $\dot{\Sigma}_{*}$ as a function of the atomic gas surface density
    $\Sigma_{\rm HI}$. Lower right: star formation rate $\dot{\Sigma}_{*}$ as a function of the total gas surface density
    $\Sigma_{\rm g}=\Sigma_{\rm H_{2}}+\Sigma_{\rm HI}$. The solid line corresponds to a star formation timescale of
    $t_{*}=2$~Gyr. The arrow takes into account the CO detection limit.
  \label{fig:plot_n4522_ha}}%
\end{figure*}

\begin{figure*}
  \centering
  \includegraphics[width=14cm]{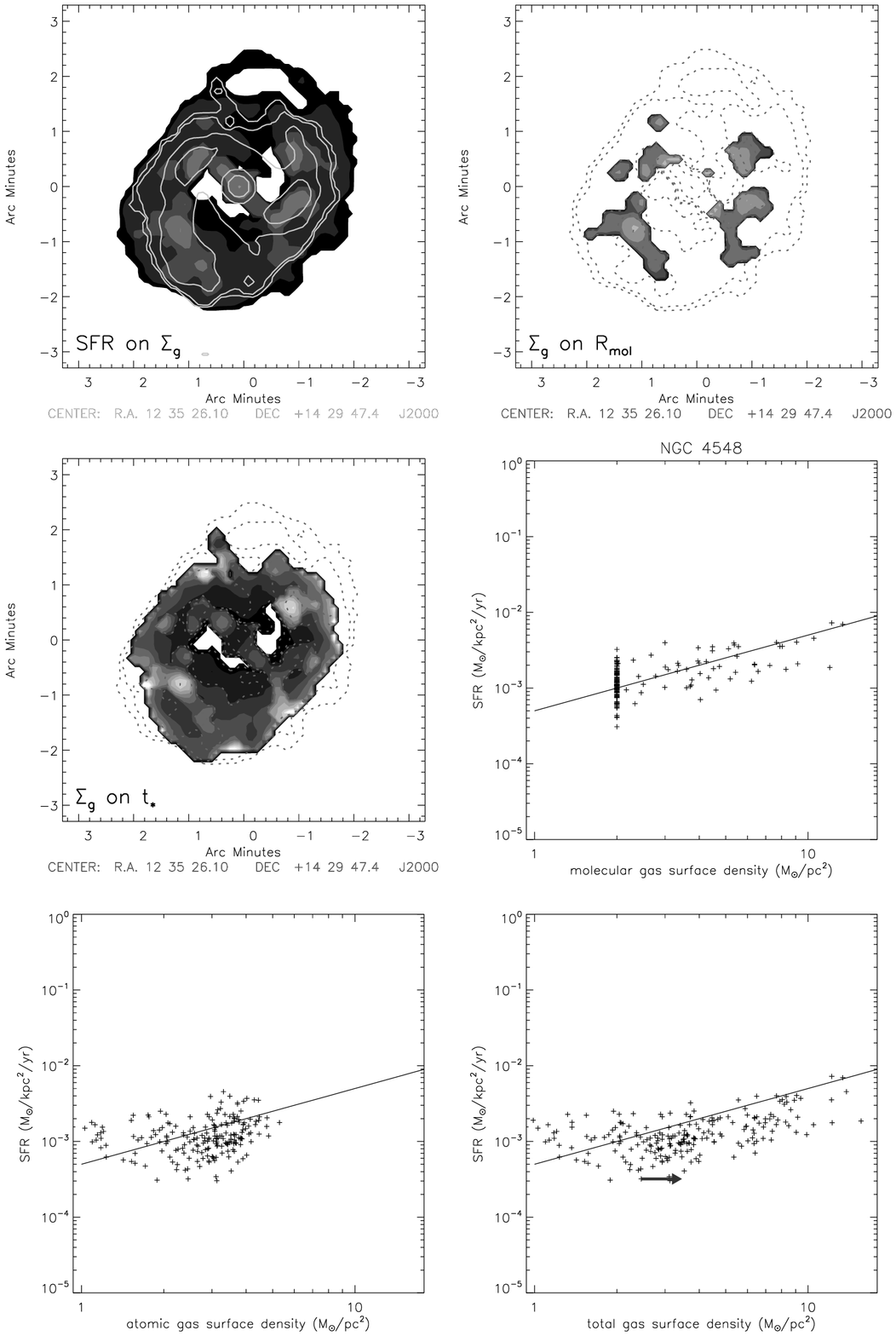}
  \caption{NGC~4548: top left: star formation rate (from UV+TIR) in contours on a greyscale map of the total gas surface density.
    Contour levels are $(1,2,4,8,16)\ 3 \times 10^{-4}$~M$_{\odot}$kpc$^{-2}$pc$^{-1}$.
    Greyscale levels are $(1,2,4,8,16)$~M$_{\odot}$pc$^{-2}$. Top right: 
    Total gas surface density contours on the molecular gas fraction $\Sigma_{\rm H_{2}}/\Sigma_{\rm HI}$.
    Greyscale levels are $(1,2,4,8,16,32) \times 0.1$.
    Middle left: total gas surface density contours on the star formation timescale $t_{*}=\Sigma_{\rm g}/\dot{\Sigma}_{*}$.
    Greyscale levels are $(1,10,20,30,40,50,60,70,80,90,100) \times 10$~Myr.
    Middle right: star formation rate $\dot{\Sigma}_{*}$ as a function of the molecular gas surface density
    $\Sigma_{\rm H_{2}}$. The vertical set of points at the lowest $\Sigma_{\rm mol}$ are upper limits for $\Sigma_{\rm mol}$.
    Lower left: star formation rate $\dot{\Sigma}_{*}$ as a function of the atomic gas surface density
    $\Sigma_{\rm HI}$. Lower right: star formation rate $\dot{\Sigma}_{*}$ as a function of the total gas surface density
    $\Sigma_{\rm g}=\Sigma_{\rm H_{2}}+\Sigma_{\rm HI}$. The solid line corresponds to a star formation timescale of
    $t_{*}=2$~Gyr. The arrow takes into account the CO detection limit.
  \label{fig:plot_n4548}}%
\end{figure*}

\begin{figure*}
  \centering
  \includegraphics[width=14cm]{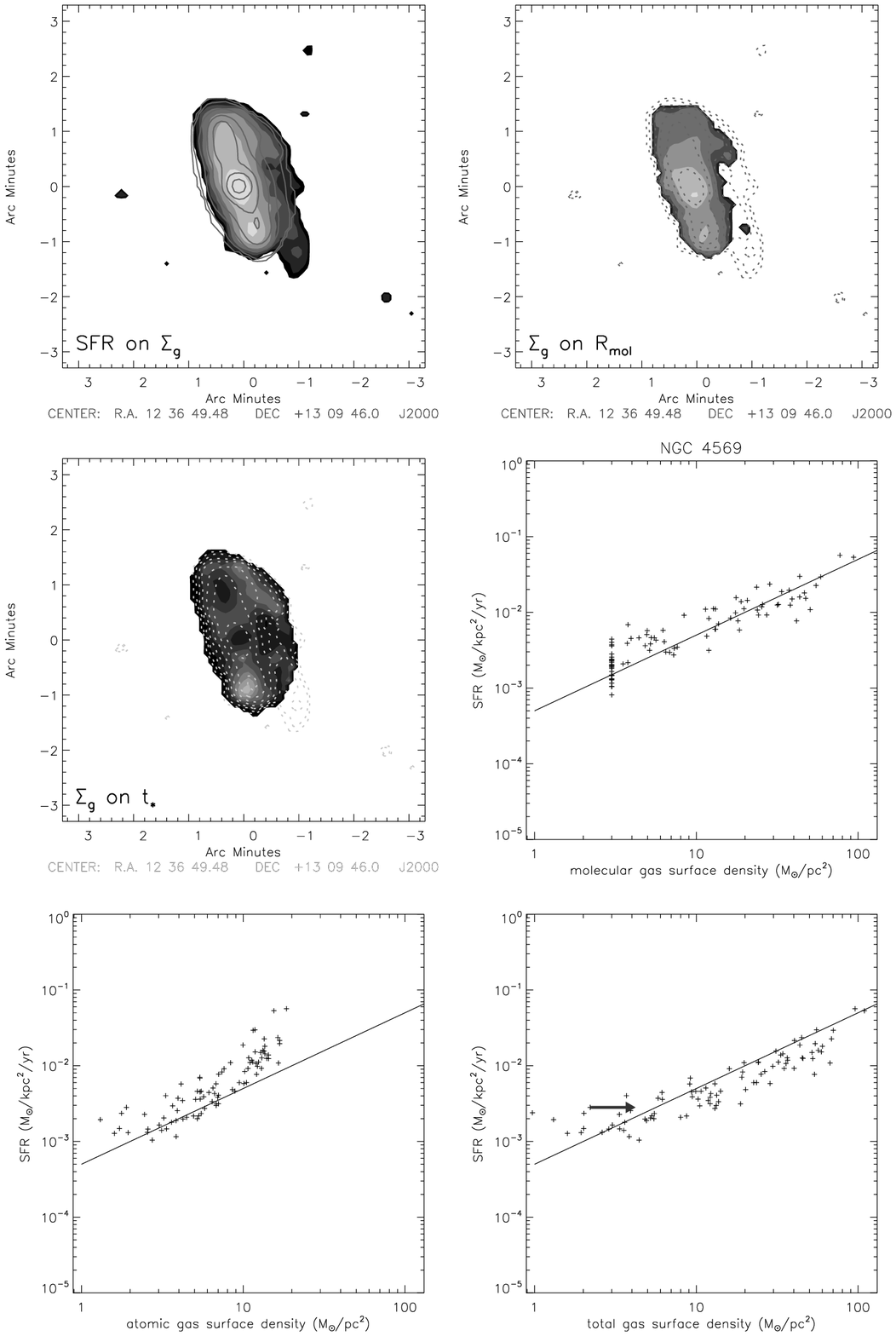}
  \caption{NGC~4569: top left: star formation rate (from UV+TIR) in contours on a greyscale map of the total gas surface density.
    Contour levels are $(1,2,4,8,16,32,64)\ 8 \times 10^{-4}$~M$_{\odot}$kpc$^{-2}$pc$^{-1}$.
    Greyscale levels are $(1,2,4,8,16,32,64)$~M$_{\odot}$pc$^{-2}$. Top right: 
    Total gas surface density contours on the molecular gas fraction $\Sigma_{\rm H_{2}}/\Sigma_{\rm HI}$.
    Greyscale levels are $(1,2,4,8,16,32,64) \times 0.1$.
    Middle left: total gas surface density contours on the star formation timescale $t_{*}=\Sigma_{\rm g}/\dot{\Sigma}_{*}$.
    Greyscale levels are $(1,10,20,30,40,50,60,70) \times 10$~Myr.
    Middle right: star formation rate $\dot{\Sigma}_{*}$ as a function of the molecular gas surface density
    $\Sigma_{\rm H_{2}}$. The vertical set of points at the lowest $\Sigma_{\rm mol}$ are upper limits for $\Sigma_{\rm mol}$.
    Lower left: star formation rate $\dot{\Sigma}_{*}$ as a function of the atomic gas surface density
    $\Sigma_{\rm HI}$. Lower right: star formation rate $\dot{\Sigma}_{*}$ as a function of the total gas surface density
    $\Sigma_{\rm g}=\Sigma_{\rm H_{2}}+\Sigma_{\rm HI}$. The solid line corresponds to a star formation timescale of
    $t_{*}=2$~Gyr. The arrow takes into account the CO detection limit.
  \label{fig:plot_n4569}}%
\end{figure*}

\begin{figure*}
  \centering
  \includegraphics[width=14cm]{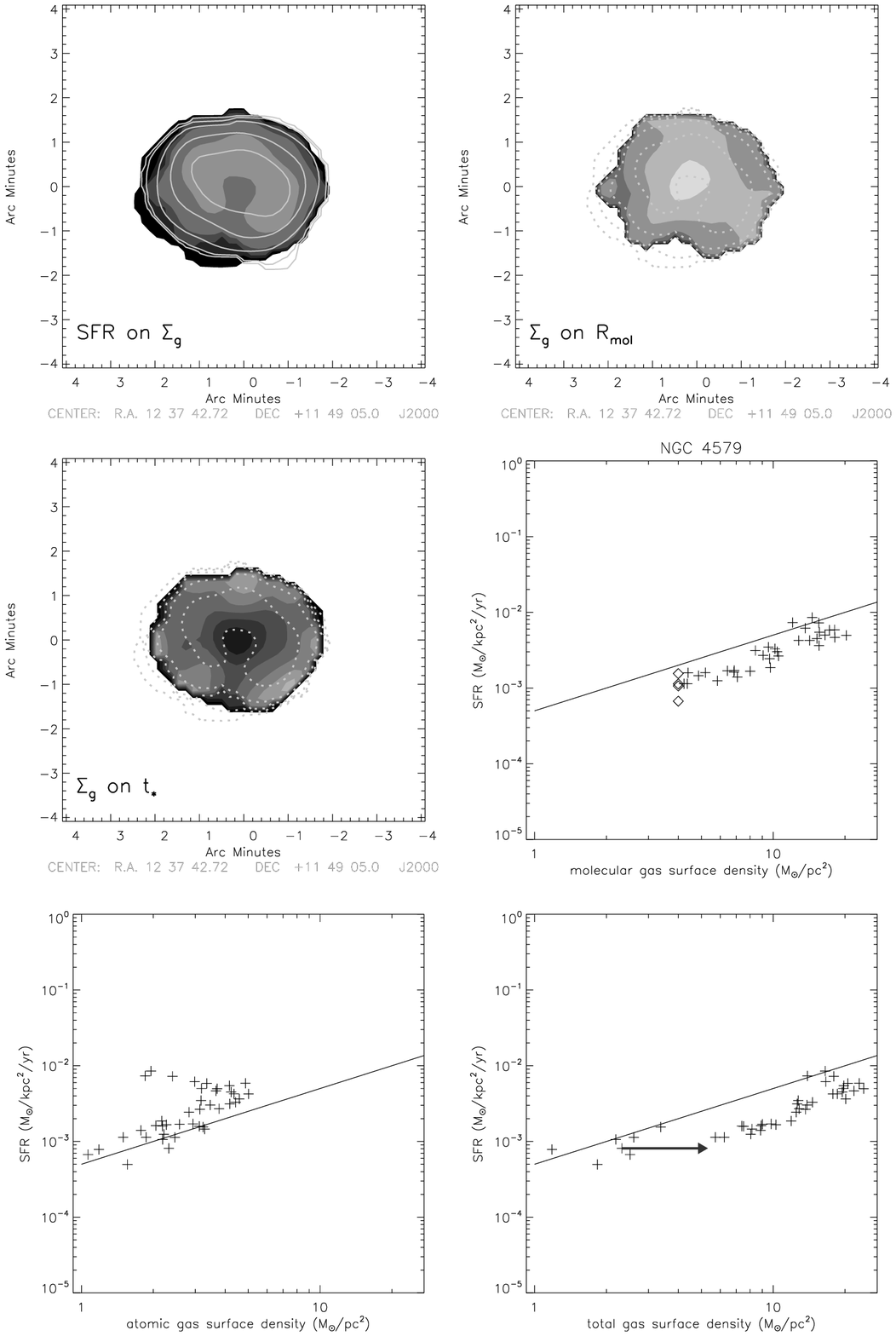}
  \caption{NGC~4579: top left: star formation rate (from UV+TIR) in contours on a greyscale map of the total gas surface density.
    Contour levels are $(1,2,4,8,16)\ 3 \times 10^{-4}$~M$_{\odot}$kpc$^{-2}$pc$^{-1}$.
    Greyscale levels are $(1,2,4,8,16)$~M$_{\odot}$pc$^{-2}$. Top right: 
    Total gas surface density contours on the molecular gas fraction $\Sigma_{\rm H_{2}}/\Sigma_{\rm HI}$.
    Greyscale levels are $(1,2,4,8,16,32,64) \times 0.1$.
    Middle left: total gas surface density contours on the star formation timescale $t_{*}=\Sigma_{\rm g}/\dot{\Sigma}_{*}$.
    Greyscale levels are $(1,10,20,30,40,50,60,70) \times 10$~Myr.
    Middle right: star formation rate $\dot{\Sigma}_{*}$ as a function of the molecular gas surface density
    $\Sigma_{\rm H_{2}}$. Diamonds are upper limits for $\Sigma_{\rm mol}$.
    Lower left: star formation rate $\dot{\Sigma}_{*}$ as a function of the atomic gas surface density
    $\Sigma_{\rm HI}$. Lower right: star formation rate $\dot{\Sigma}_{*}$ as a function of the total gas surface density
    $\Sigma_{\rm g}=\Sigma_{\rm H_{2}}+\Sigma_{\rm HI}$. The solid line corresponds to a star formation timescale of
    $t_{*}=2$~Gyr. The arrow takes into account the CO detection limit.
  \label{fig:plot_n4579}}%
\end{figure*}

\begin{figure*}
  \centering
  \includegraphics[width=14cm]{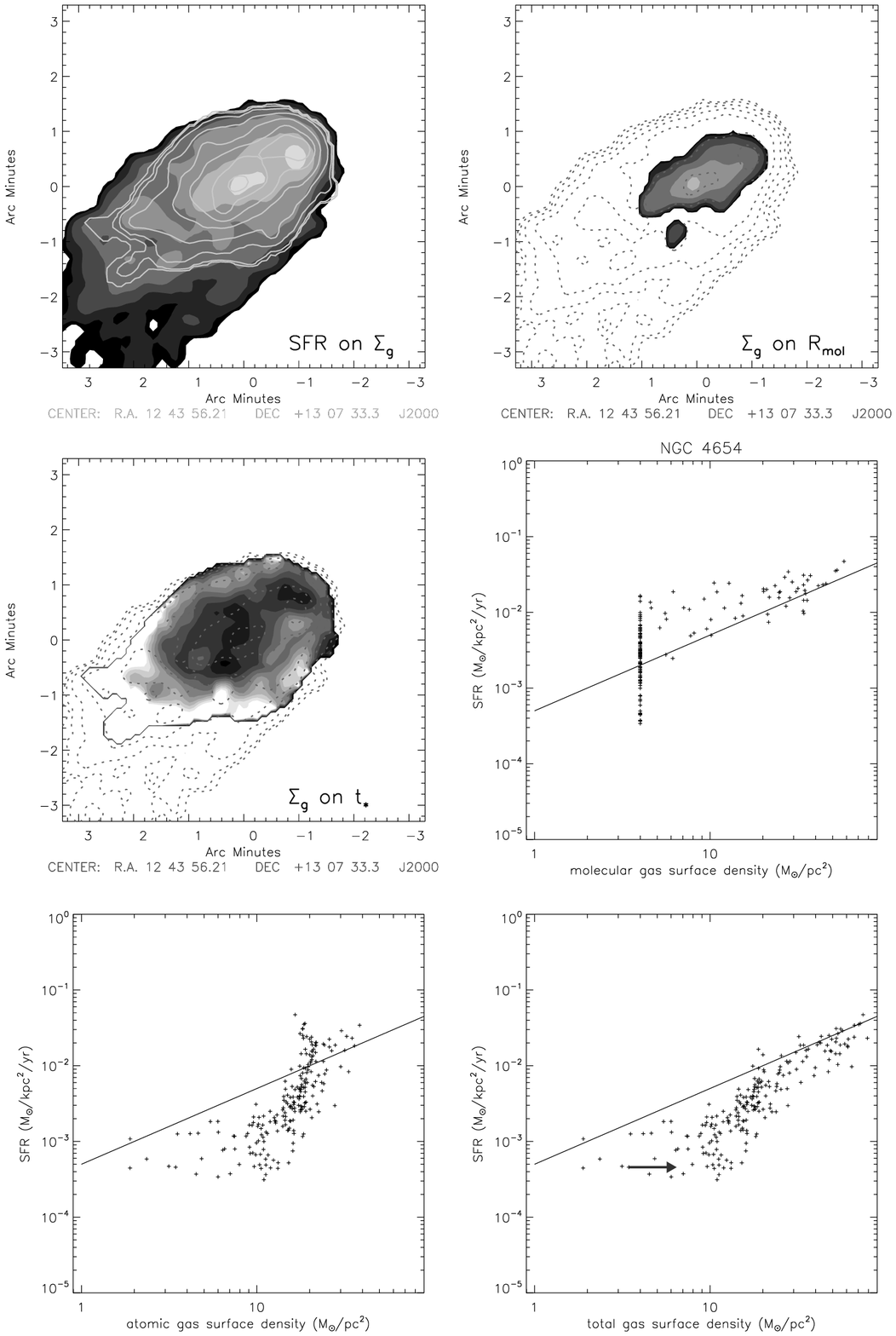}
  \caption{NGC~4654: top left: star formation rate (from UV+TIR) in contours on a greyscale map of the total gas surface density.
    Contour levels are $(1,2,4,8,16,32,64,128)\ 3 \times 10^{-4}$~M$_{\odot}$kpc$^{-2}$pc$^{-1}$.
    Greyscale levels are $(1,2,4,8,16,32,64)$~M$_{\odot}$pc$^{-2}$. Top right: 
    Total gas surface density contours on the molecular gas fraction $\Sigma_{\rm H_{2}}/\Sigma_{\rm HI}$.
    Greyscale levels are $(1,2,4,8,16,32) \times 0.1$.
    Middle left: total gas surface density contours on the star formation timescale $t_{*}=\Sigma_{\rm g}/\dot{\Sigma}_{*}$.
    Greyscale levels are $(1,10,20,30,40,50,60,70,80,90,100) \times 10$~Myr.
    Middle right: star formation rate $\dot{\Sigma}_{*}$ as a function of the molecular gas surface density
    $\Sigma_{\rm H_{2}}$. The vertical set of points at the lowest $\Sigma_{\rm mol}$ are upper limits for $\Sigma_{\rm mol}$.
    Lower left: star formation rate $\dot{\Sigma}_{*}$ as a function of the atomic gas surface density
    $\Sigma_{\rm HI}$. Lower right: star formation rate $\dot{\Sigma}_{*}$ as a function of the total gas surface density
    $\Sigma_{\rm g}=\Sigma_{\rm H_{2}}+\Sigma_{\rm HI}$. The solid line corresponds to a star formation timescale of
    $t_{*}=2$~Gyr. The arrow takes into account the CO detection limit.
  \label{fig:plot_n4654}}%
\end{figure*}

\end{document}